\documentclass[twocolumn,amssymb,amsmath,floats,showpacs,superscriptaddress,pre,floatfix]{revtex4-1}

\usepackage{amsmath,amssymb,bm}
\usepackage{graphics}
\usepackage{epsfig}
\usepackage{graphicx}

\begin{document}

\title{Critical phenomena and noise-induced phase transitions in neuronal networks}

\author{K.-E. Lee}
\affiliation{
Department of Physics $\&$ I3N, University of Aveiro,
3810-193 Aveiro, Portugal}
\author{M. A. Lopes}
\affiliation{
Department of Physics $\&$ I3N, University of Aveiro,
3810-193 Aveiro, Portugal}
\author{J. F. F. Mendes}
\affiliation{
Department of Physics $\&$ I3N, University of Aveiro,
3810-193 Aveiro, Portugal}
\author{A. V. Goltsev}
\affiliation{
Department of Physics $\&$ I3N, University of Aveiro,
3810-193 Aveiro, Portugal}
\affiliation{Ioffe Physical-Technical Institute, 194021 St.~Petersburg, Russia}

\begin{abstract}
We study numerically and analytically
first- and second-order phase transitions in neuronal networks stimulated by shot noise (a flow of random spikes bombarding neurons).
Using an exactly solvable cortical model of neuronal networks on classical random networks, we find critical phenomena accompanying the transitions and their dependence on the shot noise intensity. We show that a pattern of spontaneous neuronal activity near a critical point of a phase transition is a characteristic property that can be used to identify the bifurcation mechanism of the transition.
We demonstrate that bursts and avalanches are precursors of a first-order phase transition, paroxysmal-like spikes of activity precede a second-order phase transition caused by a saddle-node bifurcation, while irregular spindle oscillations represent spontaneous activity near a second-order phase transition caused by a supercritical Hopf bifurcation.
Our most interesting result is the observation of the paroxysmal-like spikes. We show that a paroxysmal-like spike is a single nonlinear event
that appears instantly from a low background activity with a rapid onset, reaches a large amplitude,
and ends up with an abrupt return to lower activity. These spikes are similar to single paroxysmal spikes and sharp waves observed in EEG measurements.
Our analysis shows that above the saddle-node bifurcation, sustained network oscillations appear with a large amplitude but a small frequency in contrast to network oscillations near the Hopf bifurcation that have a small amplitude but a large frequency.
We discuss an amazing similarity between excitability of the cortical model stimulated by shot noise and  excitability of the Morris-Lecar neuron stimulated by an applied current.

\end{abstract}



\maketitle

\section{Introduction}
\label{intro}

In the brain, interactions among neurons lead to diverse collective phenomena such as, for example, self-organization, phase transitions, brain rhythms, and avalanches \cite{Kelso_1995,Chialvo_2006,Buzsaki_2006}.
Among phase transitions, one can mention a non-equilibrium second-order phase transition
observed in human bimanual coordination \cite{Kelso_1984,hkb1985,Kelso_1986}.
Brain rhythms, epileptic seizures, and the ultraslow oscillations of BOLD fMRI patterns may also emerge as a result of non-equilibrium second-order phase transitions \cite{Steyn-Ross_2010}. Living neural networks stimulated by an electric field undergo a first-order phase transition that can be seen as a jump of neuronal activity at a certain applied voltage \cite{Breskin_2006}. Taking into account the role played by the collective phenomena in brain dynamics, it is very important to understand their nature and mechanisms.
It is well known that bifurcations are responsible for the emergence of oscillations in nonlinear dynamic models \cite{Strogatz_book1994,Kuznetsov_1998}, for example, the Hodgkin-Huxley model of a biological neuron \cite{Rinzel1989,Izhikevich2000}.
In the context of brain rhythms, the Hopf bifurcation was discussed in the framework of mean-field cortical models \cite{Steyn-Ross_2010}, models of randomly connected integrate-and-fire neurons \cite{amit97,bh1999,b2000,obh2009,lb2011,ldl2005,bh2008,mmkn10}, and networks of stochastic spiking neurons \cite{Benayoun_2010,Wallace_2011}.
However, when studying a phase transition,
it is not enough to identify the type of bifurcation. It is also necessary to reveal and study critical phenomena accompanying the transition
\cite{Stanley_book}. In the brain, various patterns of spontaneous activity representing collective phenomena were observed, such as  neuronal avalanches
\cite{Beggs_2003,Chialvo_2006,Plenz_2007},
paroxysmal activity \cite{Steriade_1995,Timofeev_2004}, spindle oscillations \cite{cdss1996}, and many others.
Despite a significant progress, understanding of collective phenomena in the brain and bifurcation mechanisms of phase transitions
is elusive.

A neuronal network undergoes a phase transition from one to another state when a control parameter, such as an applied voltage or a flow of spikes bombarding neurons, reaches a critical value. In some cases (for example, for epileptic seizures), it is necessary  to foresee that a neuronal network approaches to the critical point. An analysis of
patterns and statistics of spontaneous neuronal activity and critical phenomena near the critical point may be a useful method for solving the problem.
Nowadays, a comprehensive analysis of the critical phenomena in neuronal networks is far from to be complete.

In statistical physics, exactly solved models
largely help us to understand phase transitions and critical phenomena \cite{Baxter_1982}. Unfortunately, even simple versions of neuronal networks composed of
integrate-and-fire neurons are very complex for an analytical consideration
\cite{amit97,bh1999,b2000,obh2009,lb2011,ldl2005,bh2008,mmkn10}. In the present paper,
we study analytically and numerically
an exactly solvable cortical model with stochastic excitatory and inhibitory neurons on complex networks. In the framework of this model, we consider  first- and second-order phase transitions stimulated by shot noise (a flow of random spikes bombarding neurons). We also study critical phenomena accompanying the transitions and patterns of spontaneous activity signaling the transitions.
First, we study a noise-induced first-order phase transition from low to high neuronal activity. The transition occurs if
inhibitory neurons respond faster on stimuli than excitatory neurons.
We demonstrate that bursts and avalanches of neuronal activity
precede the transition.
Second, we study two noise-induced second-order phase transitions that occur if inhibitory neurons respond slower on stimuli than excitatory neurons.
The transitions represent two scenarios of appearance and disappearance of sustained network oscillations.
We show that, when increasing the shot noise intensity, at first, sustained network oscillations appear due to a saddle-node bifurcation
and then, at a higher shot noise intensity, the oscillations disappear due to a supercritical Hopf bifurcation.
We study patterns of spontaneous neuronal activity near
the bifurcations. We show that sharp paroxysmal-like spikes of activity precede the second-order phase transition caused by the saddle-node bifurcation.
Above the Hopf bifurcation, spontaneous activity appears in a form of irregular spindles formed by damped oscillations. We also study analytically and numerically
sustained network oscillations near the critical points of the bifurcations.
Furthermore, we analyze the power spectral density (PSD) of spontaneous neuronal activity and its dependence on the noise intensity.
We show that the PSD depends strongly on the bifurcation mechanism and the closeness to the critical point. We compare our results with experimental data and previous theoretical investigations. Finally, we discuss an amazing similarity between excitability of the considered cortical model stimulated by shot noise and  excitability of the Morris-Lecar neuron \cite{ML81} stimulated by an applied current.

\section{Model}
\label{Model}

We study a cortical model
composed of $N_e$ excitatory and $N_i$ inhibitory neurons. $N_e+N_i\equiv N$ is the network size, $g_{e(i)}\equiv N_{e(i)}/N$ is the fraction of excitatory (inhibitory) neurons. Neurons are randomly connected with probability $c/N$ and form a classical random graph with Poisson degree distribution and the mean degree $c$. The network is locally tree-like and has the small-world properties \cite{Albert_2002,dg2002,newman2003} similar to ones found in brain networks \cite{sporns04}.
Our model also takes into account noise playing an important role in the brain dynamics \cite{white00,lgns04,faisal08,ermentrout08}. We assume that
neurons are bombarded by random spikes represented by Dirac delta functions,
\begin{equation}
I(t)=\sum_{i} q \delta(t-t_i),
\label{s-n}
\end{equation}
where $t_i$ are arrival times of spikes and $q$ is their amplitude. This kind of random input is so-called shot noise. The flow of random spikes bombarding neurons represents a combined effect of synaptic noise (spontaneous release of neurotransmitters), stimuli from other brain areas or sensory stimuli.
According to Schottky's result, in the case of the Poisson distribution of
interspike intervals, the power spectral density $S(\omega)$
is proportional to the mean frequency $\omega_{sn}$ of spikes, $S(\omega)=2q^2 \omega_{sn}$.
In the present paper, we assume that the probability to receive $\xi$ random spikes during the integration
time $\tau$ is Gaussian,
\begin{equation}
G(\xi)=G_0 \exp[-(\xi-\langle n\rangle)^2 / 2\sigma^2],
\label{gaussian}
\end{equation}
where $\sigma^2$ is the variance, $\langle n\rangle= \omega_{sn}\tau$ is the mean number of spikes arriving during the time interval $\tau$, and $G_0$ is the normalization constant, $\sum_{\xi=0}^{\infty}G(\xi)=1$. We use $\langle n\rangle$ as the control parameter characterizing the shot noise intensity.

Neurons also receive delta-like
spikes from active neighbors. The spikes mediate interaction among neurons. We assume that efficacies of synaptic connections with excitatory and inhibitory neurons are uniform and equal to $J_e$ ($J_e >0$) and $J_i$ ($J_i <0$), respectively.
The total input $I(t)$ includes spikes from
shot noise and excitatory and inhibitory presynaptic neurons.
We define the input $V_n$ to a neuron with index $n$, $n=1,2,\dots N$, as the integral of $I(t)$ over the time interval $[t-\tau,t]$. It gives
\begin{equation}
V_n(t)= \xi q + kJ_e +lJ_i,
\label{input}
\end{equation}
where $\xi$, $k$, and $l$ are the numbers of spikes arriving during the time interval $[t-\tau,t]$ from shot noise, active presynaptic excitatory and inhibitory neurons, respectively. The numbers $k$ and $l$ are random and are determined by activity of presynaptic neurons during the interval $[t-\tau,t]$. The network structure is encoded in the adjacency matrix.

In our model, neurons are tonic and the firing frequency $f(V)$ versus input $V$ is the Heaviside function
\begin{equation}
f(V)=f \Theta(V-V_{th}),
\label{f-I}
\end{equation}
where $V_{th}$ is a threshold. The frequency
$f$ is the same for both excitatory and inhibitory neurons. If $f\tau <1$ and spike emission times of neurons are uncorrelated, then
during the time interval $[t-\tau,t]$, each active presynaptic neuron contributes to $V_n(t)$ either one spike with probability $\tau f$ or none with probability $1-\tau f$.

We consider stochastic neurons like those of \cite{Benayoun_2010,Wallace_2011,goltsev10,goltsev13}.
It means that the response of a neuron to an input is a stochastic process. Such a stochastic behavior might be caused by cellular noise and intensive bombardment by random spikes.

Two rules determines dynamics of the cortical model:
\begin{enumerate}
\item If the input $V_n(t)$ at an inactive excitatory (inhibitory) neuron $n$ at time $t$  is at least a certain threshold $V_{th}$, then this
neuron is activated with probability $\mu_{e} \tau$ ($\mu_{i} \tau$) and fires spikes.
\item An active excitatory (inhibitory) neuron $n$ is inactivated with probability $\mu_{e}\tau$ ($\mu_{i}\tau$) if $V_n(t)< V_{th}$.
\end{enumerate}
We introduce a dimensionless activation threshold $\Omega \equiv V_{th}/J_{e} $.
$\Omega$ is of the order of 15-30 in living neuronal networks
\cite{Eckmann07,breskin06,Soriano08}
and about $30-400$ in the brain.
In our model, $1/\mu_{e}$ and $1/\mu_{i}$ are of the order of the first-spike latencies of excitatory and inhibitory neurons (from 6 to 100 ms in the cortex \cite{Heil04,Swadlow03,fmi04,mfmt05}).
%
%
We introduce a parameter,
\begin{equation}
\alpha \equiv \mu_{i}/\mu_{e}.
\label{alpha}
\end{equation}
If inhibitory neurons respond faster on stimuli than inhibitory neurons, i.e., the response time $T_i=1/\mu_{i}$ of an inhibitory neuron is smaller than the response time $T_e=1/\mu_{e}$ of an excitatory neuron,  then $\alpha >1$.
If excitatory neurons respond faster,
i.e., $T_e <T_i$, then $\alpha <1$.
In the cortex,
$\alpha$ may be both larger and smaller than 1 \cite{Heil04,Swadlow03,fmi04,mfmt05}.


\subsection{Rate equations}
\label{Rate equations}

The behavior of the cortical model
is described by the fractions $\rho_e(t)$ and
$\rho_i(t)$ of active
excitatory and inhibitory neurons, respectively, at time $t$. We will call them
`activities'.
We assume that activities are changed slightly during the integration time $\tau$.
Using the rules
formulated above and the methods developed in \cite{goltsev10,goltsev13,dorogovtsev08}, in particular, the method of generating functions \cite{goltsev13}, in the limit $N\rightarrow\infty$, we find explicit rate equations,
\begin{eqnarray}
& \dot{\rho}_e(t)=F_{e}(t)[1{-}\rho_e(t)]-\mu_e \rho_{e}(t)+\mu_e \Psi_{e}(\rho_{e}(t),\rho_{i}(t)), \nonumber \\
& \!\!\!\! \dot{\rho}_i(t)=F_{i}(t)[1{-}\rho_i(t)]-\mu_i \rho_{i}(t) + \mu_i \Psi_{i}(\rho_{e}(t),\rho_{i}(t)),
\label{eq:10}
\end{eqnarray}
where $\dot{\rho}\equiv d \rho/dt$. $\Psi_{e(i)}(\rho_e,\rho_i)$ is the probability
that, at given activities $\rho_e$ and $\rho_i$, input to a randomly chosen excitatory (inhibitory) neuron is at least  $\Omega$. $F_{e}$ and $F_{i}$ represent
fields acting on excitatory and inhibitory neurons. Note that the rate equations (\ref{eq:10}) are similar to the Wilson-Cowan equations \cite{Wilson_1972} and rate equations derived for a stochastic rate model in \cite{Benayoun_2010,Wallace_2011}. In the case of the classical random graph, we find
\begin{eqnarray}
& \Psi_i(\rho_e,\rho_i)=\Psi_e(\rho_e,\rho_i)\equiv
\Psi(\rho_e,\rho_i) = \nonumber \\
&\!\!\!\!\!\!\!\!\!\sum_{k,l,\xi=0}^{\infty}\Theta(k J_{e}{+}lJ_{i}{+}\xi q{-}\Omega J_{e}) G(\xi)
P_{k}(g_{e}\rho_{e} \widetilde{c})P_{l}(g_{i}\rho_{i} \widetilde{c}),
\label{eq:14}
\end{eqnarray}
where $\widetilde{c} \equiv c\tau f$. $G(\xi)$,
$P_k(g_e\rho_e \widetilde{c})$, and $P_l(g_i\rho_i \widetilde{c})$ are the probabilities that,
during the time interval $\tau$,
a randomly chosen neuron receives $\xi$ random spikes from shot noise, $k$ spikes from excitatory neurons, and $l$ spikes from inhibitory neurons, respectively. Note that the Poisson function $P_k(c)\equiv c^k e^{-c}/k!$ is the probability that a randomly chosen neuron has $k$ presynaptic connections. Below we will study analytically and numerically  Eqs.~(\ref{eq:10}) and compare with simulations of the cortical model.


Our cortical model based on \cite{goltsev10} is similar to the stochastic model of spiking neurons proposed by Benayoun \emph{et al.} \cite{Benayoun_2010}.
Both models consider networks of stochastic neurons (`input-dependent stochastic switches' by \cite{Benayoun_2010}). The difference between the models is in some details about how to describe activation and deactivation processes and external input. Benayoun \emph{et al.} \cite{Benayoun_2010} assume that each neuron spikes with a 
rate dependent on its total synaptic input, while the resulting spiking activity decays at a constant rate independent on the input. In our model, we use a similar activation rule, while spiking activity decays with a certain rate only if the input becomes smaller than a threshold. The rates for activation and decay are different in \cite{Benayoun_2010}, in contrast to our model where they are the same. Benayoun \emph{et al.} assume that external input to each neuron is fixed in contrast to our model where external input is represented by shot noise. It is not surprising that, despite these differences, these models
demonstrate similar dynamics.
The advantage of the models with stochastic neurons is that they can be solved explicitly. Benayoun \emph{et al.} \cite{Benayoun_2010} and Wallace \emph{et al.} \cite{Wallace_2011} derived explicit rate equations for networks with all-to-all connections while sparse randomly connected networks (classical random networks) were studied numerically.
Methods of complex network theory \cite{dorogovtsev08} allowed us to find explicit rate equations for neuronal networks on classical random graphs \cite{goltsev13} and scale-free networks \cite{goltsev13} and apply the model to study stochastic resonance \cite{Lopes2012} and the role of synaptic plasticity  \cite{Lee2012}.

\subsection{Algorithm of simulations and parameters}
\label{Algorithm}

In simulations,
we built a directed network, linking neurons with the probability $c/N$.
We divided time into intervals of width $\Delta t=\tau$. At each time step, for each neuron we calculated input Eq.~(\ref{input})
given that each active presynaptic neuron contributes a spike with probability $\tau f$. The number of random spikes (shot noise) in this input was generated
by the Gaussian process $G(\xi)$.
Then, we updated states of neurons, using the rules formulated above. In our paper, we present numerical calculations for parameters $N=10^5$, $c=10^3$, $\Omega=30$, $\tau f = 0.1$, $f=\mu_{e}$, and $g_i=0.25$.
We analyze dynamical behavior of the cortical model in dependence on only two parameters: the parameter $\alpha$ and the shot noise intensity. The latter parameter is the control parameter.
Throughout this paper we use $1/\mu_{e}\equiv 1$ as time unit and $J_e\equiv 1$ as input unit. Following \cite{amit97}, we choose $J_{i}=-3J_{e}$. We use $q=J_{e}$ and $\sigma^2 =10$ for the amplitude and variance of shot noise.

\subsection{Steady states}
\label{steady states}

\begin{figure}
\includegraphics[width=0.25\textwidth]{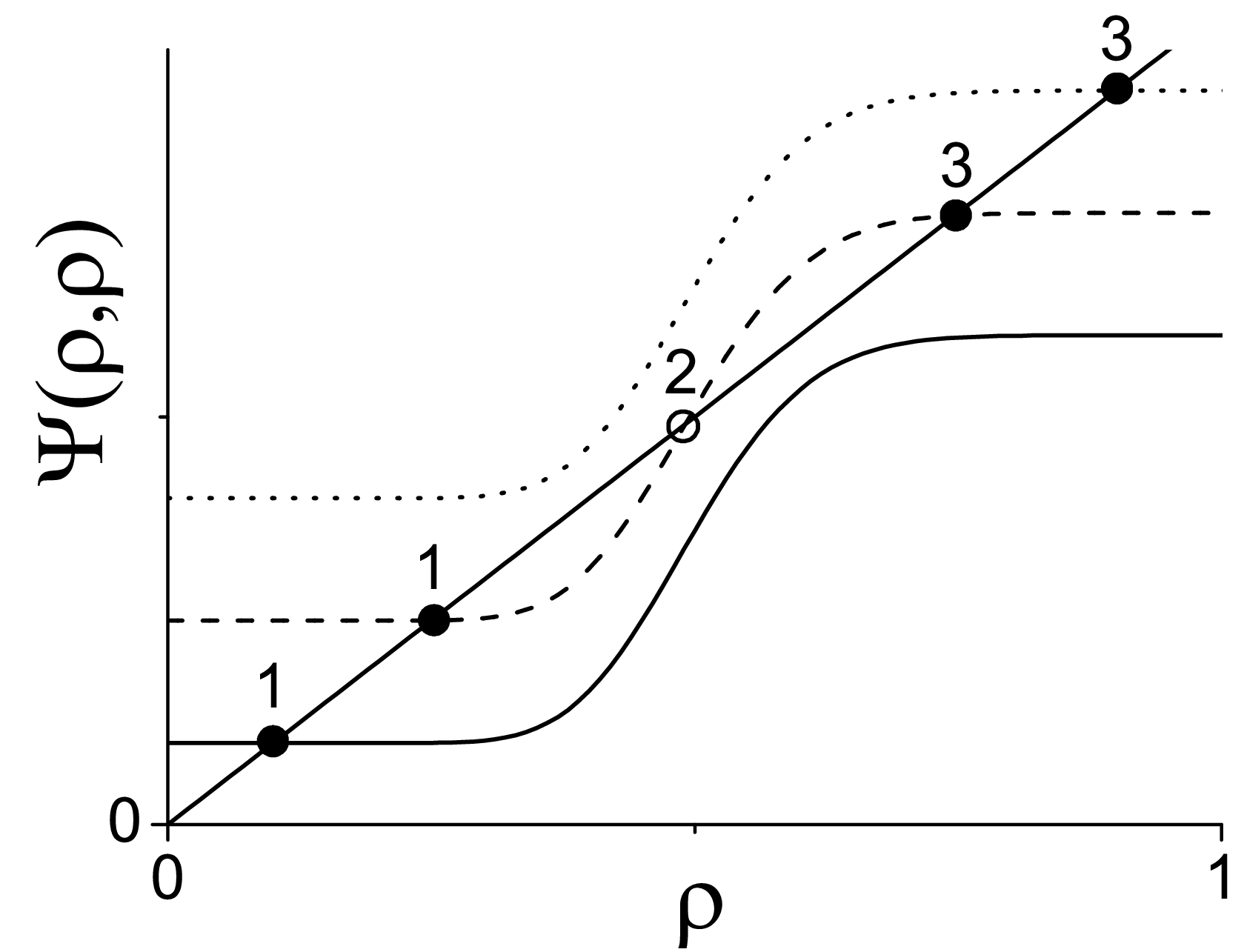}
\caption{Points 1,2, and 3 represent solutions of the steady state equation $\rho=\Psi(\rho,\rho)$ for the cases  $\langle n \rangle <n_{c1}$ (solid line),  $n_{c1} < \langle n \rangle <n_{c2}$ (dashed line), and $\langle n \rangle > n_{c2}$ (dotted line).
\label{fixpoints}}
\end{figure}

The shot noise intensity $\langle n \rangle$ determines activities $\rho_e$ and $\rho_i$ of excitatory and inhibitory populations at given model parameters.
At
zero fields $F_{e}=F_{i}=0$,
from Eqs.~(\ref{eq:10}) we obtain
$\rho_e =\rho_i\equiv\rho$ in a steady state ($d \rho_a /dt=0$). $\rho$ is a solution of the steady state equation, 
\begin{equation}
\rho=\Psi(\rho,\rho).
\label{steady-1}
\end{equation}
A graphical solution of the equation is displayed in Fig. \ref{fixpoints}. If the shot noise intensity $\langle n \rangle$ is either sufficiently small or large, then there is only one solution, either point 1 or point 3. These fixed points correspond to steady states with low and high neuronal activity, respectively. In an intermediate range $n_{c1} < \langle n \rangle <n_{c2}$ there are three fixed points (1,2, and 3). The critical point  $\langle n \rangle = n_{c1}$ is the point where fixed points 2 and 3 coalesce.  Fixed points 1 and 2 coalesce at $\langle n \rangle = n_{c2}$. From Fig.~\ref{fixpoints} one sees that the coalescence occurs when
\begin{equation}
d \Psi(\rho,\rho)/d\rho=1.
\label{steady-2}
\end{equation}
Together with the steady state equation (\ref{steady-1}),
the condition (\ref{steady-2}) determines the critical points $n_{c1}$ and $n_{c2}$.

While the fixed points depend on $\langle n \rangle$, but not on $\alpha$, their local stability with respect to small perturbations depends on both $\langle n \rangle$ and $\alpha$. It is determined by eigenvalues of the Jacobian of Eqs. (\ref{eq:10}),
\begin{equation}
\widehat{J}(\rho) =
\begin{pmatrix}
-1 +\partial \Psi/\partial \rho_e & \partial \Psi/\partial \rho_i \\
\alpha \partial \Psi/\partial \rho_e & -\alpha +\alpha \partial \Psi/\partial \rho_i
\end{pmatrix}
,
\label{Jacobian}
\end{equation}
calculated at the fixed points. The eigenvalues are
\begin{equation}
\lambda_{\pm}=-\frac{1}{2}(J_{11}+J_{22})\pm 
\frac{1}{2}\sqrt{(J_{11}-J_{22})^2+4J_{12}J_{21}},
\label{eigenv}
\end{equation}
where $J_{ij}$ are the entries of the Jacobian. If $\lambda_{\pm} <0$ at a fixed point, then this point is stable (attractor). If $\lambda_{\pm} > 0$, then the point is unstable.  If one of the eigenvalues $\lambda_{\pm}$ is positive and the other is negative, then the point is saddle. If Re$\lambda_{\pm} <0$ and Im$\lambda_{\pm} \neq 0$, the point is a stable spiral. If Re$\lambda_{\pm} > 0$ and Im$\lambda_{\pm} \neq 0$, the fixed point is an unstable spiral. The fixed points and their stability
determine a phase portrait of Eqs.~(\ref{eq:10}).

If the neuronal network is weakly perturbed from an equilibrium state  corresponding to a stable fixed point $\rho$, then the real and imaginary parts of $\lambda_{+}$ at this point determines a relaxation rate $\gamma_r$ to the state,
\begin{equation}
\gamma_r=-\text{Re}\lambda_{+}(\rho),
\label{r-rate}
\end{equation}
and the angular frequency $\gamma_i$ of damped oscillations about the fixed point,
\begin{equation}
\gamma_i=\text{Im}\lambda_{+}(\rho).
\label{d-freq}
\end{equation}

\begin{table}[t]
\caption{
Local stability of the fixed points 1, 2, and 3 in the regions Ia--IIIb on the phase diagram in Fig. \ref{fig-overview}, where s=stable, sd=saddle, u=unstable, sp=spiral, lc=limit cycle.}
\begin{center}
    \begin{tabular}{| c | c | c | c | c | c | c | c | c | c |}
    \hline
      & Ia   & Ib & Ic & Id      & Ie  & IIa  & IIb  & IIIa & IIIb   \\ \hline
    1 &  s   & s  & s  &  s      & s   &--    & --   & --   &  --      \\ \hline
    2 & --   & sd & sd & sd      & sd  &--    & --  & --   &  --      \\ \hline
    3 & --   & s  & s sp & u sp  & u &  s   & s sp & u sp \& lc & u \& lc \\ \hline
    \end{tabular}
\end{center}
\label{table1}
\end{table}

\begin{figure}
\includegraphics[width=0.45 \textwidth]{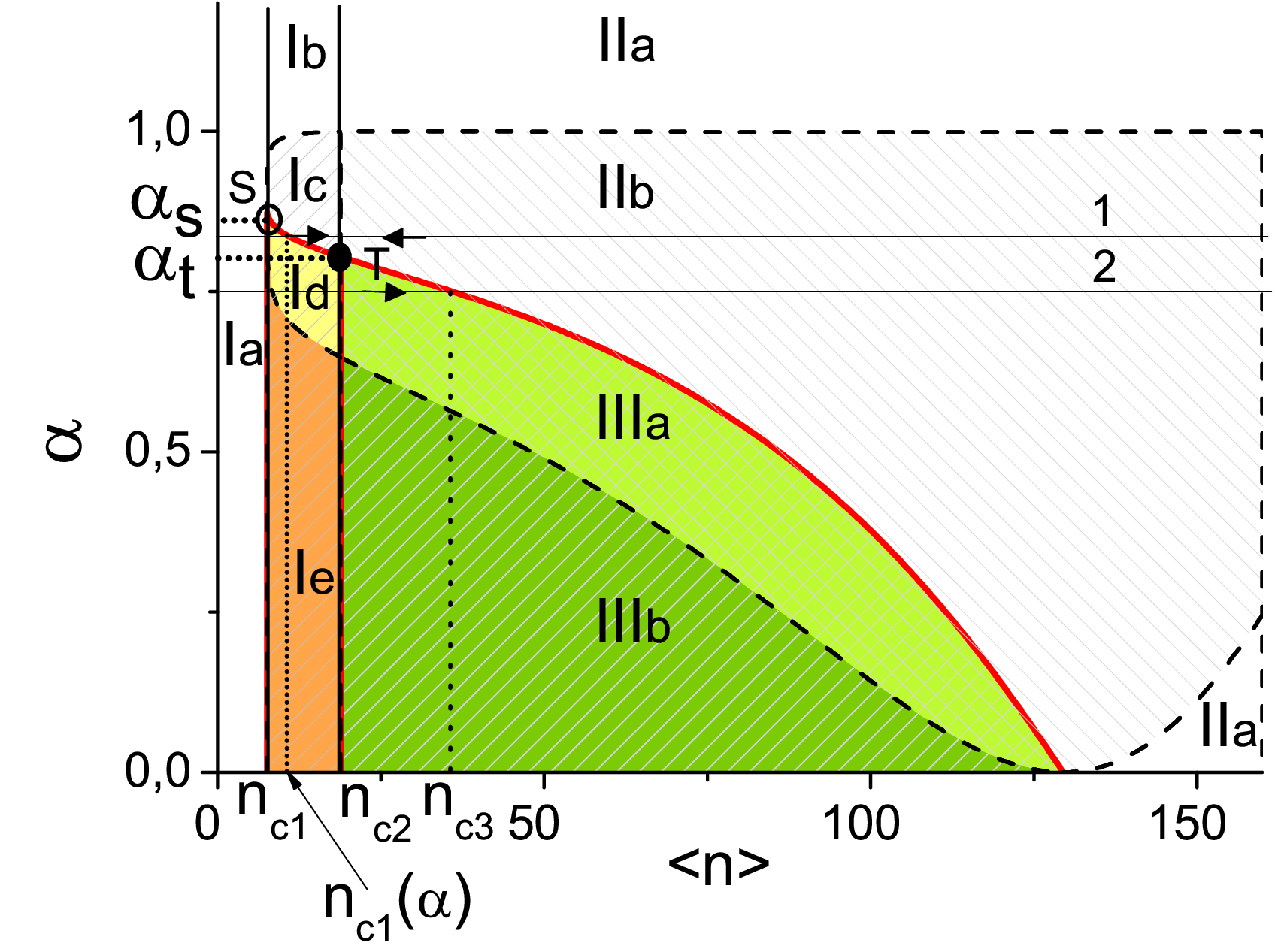}
\caption{(Color online) $\langle n \rangle -\alpha$ plane of the  phase diagram
of the cortical model.
$\langle n \rangle$ is the shot noise intensity. $\alpha$ is the ratio of the response time of excitatory neurons to the response time of inhibitory neurons. The phase regions,
the phase boundaries, and the parameters used in numerical calculations are explained in the text. The black dot represents the tricritical point \emph{T} with the coordinate $\alpha_t\approx 0.80$. Lines 1 and 2 represent two scenarios discussed in the text.
\label{fig-overview}}
\end{figure}

\subsection{Phase diagram}
\label{Phase diagram}

Analyzing the local stability of the fixed points
1, 2, and 3 in the $\alpha-\langle n \rangle$ plane (see Table \ref{table1}),
we find the phase diagram of the cortical model  displayed in Fig.~\ref{fig-overview}.
According to Table \ref{table1}, in regions Ia-Ie, the network relaxes exponentially to the stable fixed point 1 (of course, if a perturbation is small). In regions Ib and IIa, relaxation to the stable fixed point 3 is exponential while, in regions Ic and IIb, the relaxation occurs in the form of damped oscillations about the fixed point 3. In regions IIIa and IIIb, the fixed point 3 is an unstable point surrounded by a limit cycle. These are the regions with sustained network oscillations about the point 3. Nonlinear Eqs. (\ref{eq:10}) have different phase portraits in phase regions Ia--IIIb in Fig.~\ref{fig-overview}. The phase portraits in the $(\rho_e ,\rho_i)$-phase  can be found by use of the standard methods \cite{Strogatz_book1994,Kuznetsov_1998}. They determine the patterns of collective neuronal activity and response of the network on stimuli.

In Fig.~\ref{fig-overview}, the phase boundaries
are represented by the dashed and solid lines.
The vertical lines $\langle n\rangle=n_{c1}$ and $\langle n\rangle=n_{c2}$ are determined by the self-consistent solutions of Eqs. (\ref{steady-1}) and (\ref{steady-2}) discussed in Sec. \ref{steady states}.
The boundary between regions IIa and IIb and regions IIIa and IIIb is determined by the condition,
\begin{equation}
\gamma_i(\rho^{(3)})=\text{Im}\lambda_{+}(\rho^{(3)})=0
\label{pb-1}
\end{equation}
(see the dashed lines in Fig.~\ref{fig-overview}). The phase boundary between regions Ic and Id and regions IIb and IIIa is determined by the condition,
\begin{equation}
\gamma_r(\rho^{(3)})=-\text{Re}\lambda_{+}(\rho^{(3)})=0.
\label{pb-2}
\end{equation}
(see solid line in Fig.~\ref{fig-overview}). According to Eq.~(\ref{pb-2}), on the boundary between regions IIb and IIIa, the relaxation rate is zero, i.e., critical slowing down occurs.
The point $T=(n_{c2},\alpha_t)$ in Fig.~\ref{fig-overview} is a tricritical point of coexistence of three phases: the low activity state (regions Ic and Id), the high activity state (region IIb),  and the state with sustained network oscillations (region IIIa).
At the point $T$, the line of the first-order phase transition meets the lines of two continuous phase transitions.
The point $S=(n_{c1},\alpha_s)$ is the common point of regions Ia, Ic, and Id.
For the parameters used in our paper, we find $n_{c1}\approx 7.6$, $n_{c2}\approx 18.8$, $\alpha_s \approx 0.87$, and $\alpha_t \approx 0.80$.

\section{First-order phase transition}
\label{First-order phase transition}

In this section, we study critical phenomena accompanying the first-order phase transition stimulated by shot noise. In particular, we study neuronal bursts and avalanches
as precursors of the transition. Though bursts and avalanches have been broadly studied both experimentally and theoretically,
understanding of their mechanism in the brain is elusive \cite{Beggs_2003,Chialvo_2006,Plenz_2007,mmkn10,Orlandi_2013}.
Here, apart the standard measurements of the distribution function of avalanches over size, we also study critical behavior of the relaxation rate, a dependence of the power spectral density (PSD) of activity fluctuations on the shot noise intensity, and discuss finite-size effects. We find a dramatic increase of the zero frequency peak of the PSD when the shot noise intensity tends to a critical point while above the point the relaxation rate is non-zero and there are no critical fluctuations.

\begin{figure}
\includegraphics[width=0.45 \textwidth]{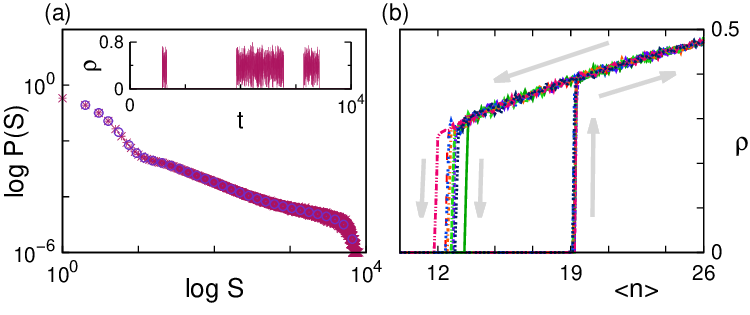}
\caption{(Color online)  (a)
Avalanche size distribution $P(s)$ versus size $s$ at $\langle n\rangle=18.8$ found by use of simulations. Inset: temporal activity of excitatory neurons
near the first-order phase transition. Time $t$ is in units of $1/\mu_e$. (b) Hysteresis in neuronal activity for increasing and decreasing noise level $\langle n\rangle$. In simulations, $\alpha=0.85$.
\label{fig-hysteresis}}
\end{figure}

The first-order phase transition occurs if
$\alpha >\alpha_t$, i.e., when the response time $T_i$ of an inhibitory neuron to stimulus is small enough in comparison with the response time $T_e$ of an excitatory neuron.
In simulations and numerical solution of Eqs.~(\ref{eq:10}), we increased the noise level $\langle n \rangle$ from zero (region Ia) to a value in region IIa (or IIb) above the critical point $n_{c2}$ and afterwards decreased it again to a value below $n_{c2}$  (see line 1 in Fig.~\ref{fig-overview}).
When increasing the noise intensity $\langle n \rangle$, the
neuronal activity undergoes a jump at $\langle n \rangle = n_{c2}$
($ n_{c2} \approx 18.8$ in Fig.~\ref{fig-hysteresis}(b)). Therefore, the critical point $n_{c2}$ is the limiting point of the first-order phase transition. This phase transition is caused by a saddle-node bifurcation that corresponds to coalescence of the stable point 1 and the saddle point 2. Simultaneously, at $\langle n \rangle = n_{c2}$, the eigenvalue $\lambda_{+}(\rho^{(1)})$ becomes zero while $\lambda_{-}(\rho^{(1)})$ remains to be negative.
The first-order phase transition was also found in \cite{Benayoun_2010}.
The line of the first-order phase transition ends up at the point $(\alpha_t,n_{c2})$ on the phase diagram. If $\alpha < \alpha_t$, the neuronal networks undergoes a second-order phase transition at $\langle n \rangle = n_{c2}$ that will be discussed in Sec.~\ref{Non-equilibrium phase transitions}.

\subsection{Avalanches}
\label{Avalanches}

In simulations, at
$\langle n\rangle \leq n_{c2}$, we observe bursts of neuronal activity (see
Fig.~\ref{fig-hysteresis}(a)). When $\langle n\rangle \rightarrow n_{c2}$
the mean interburst interval decreases while the mean burst duration increases.
The bursts  are caused by  avalanches (activation of a single neuron triggers activation of a cluster of neurons).
These activation processes are stochastic. In our model, in networks of finite size, bursts are generated by finite-size fluctuations.
We studied avalanches,
analyzing spike time series by use of the standard method (see \cite{Beggs_2003} or the recent work \cite{Friedman_2012}).
The avalanche size distribution $P(s)$ is represented in Fig. \ref{fig-hysteresis}(a). When
$\langle n\rangle$ is close to $n_{c2}$,  $P(s)$ is powerlaw, $P(s)\propto s^{-z}$, in a broad range of $s$.
Using
the maximum likelihood estimate \citep{Touboul_2010}, we found $z \approx 1.53$ and the corresponding p-value is $p=0.89$  (the closeness of $p$
to 1 shows that the fit is good). Our estimation is close to the value 1.62 found
in a stochastic rate model \cite{Benayoun_2010}. Avalanches with the exponent $z $ about 1.5 were also found near a saddle-node bifurcation in networks of leaky integrate-and-fire neurons with short-term synaptic depression \cite{mmkn10}.
Our estimation also agrees with experimental data \cite{Beggs_2003,Friedman_2012} and the standard mean-field exponent $z=3/2$ obtained for other exactly solved models \cite{goltsev10,Sethna_1993,Sethna_2001,Goltsev_2006,Dorogovtsev_2006,Baxter_2012}.

\subsection{Hysteresis}
\label{Hysteresis}

At a given $\alpha >\alpha_t $, if $\langle n \rangle$ decreases from a value
above $n_{c2}$ to a value below $n_{c2}$,
the network activity remains as high as it was above $n_{c2}$ (see Fig. \ref{fig-hysteresis}(b)).
The activity falls to a low value only at a critical intensity $\langle n \rangle = n_{c1}(\alpha)$, which, in the general case, depends on $\alpha$ and $n_{c1} \leq n_{c1}(\alpha) \leq n_{c2}$ (see Fig.~\ref{fig-overview}). If $\alpha >\alpha_s $, where $\alpha_s$ is the $\alpha$-coordinate of the point $S$ on Fig.~\ref{fig-overview},
hysteresis  occurs in the range $n_{c1} < \langle n \rangle <n_{c2}$.
If $\alpha_t <\alpha <\alpha_s $, hysteresis  occurs in a smaller range of shot noise intensity, $n_{c1}(\alpha) < \langle n \rangle <n_{c2}$ where $n_{c1}(\alpha)$ is $\langle n \rangle$ -coordinate of the interception point of the line 1 with the phase boundary between region Ic and Id ending up at points \emph{S} and \emph{T}  on the phase diagram in Fig.~\ref{fig-overview}.  The width of the hysteresis region, e.i., $n_{c2}-n_{c1}(\alpha)$, tends to zero when $\alpha \rightarrow \alpha_t$.
At $\alpha < \alpha_t $, hysteresis is absent because, in regions Id and Ie, the fixed point 3 is unstable and there is only one stable fixed point 1. One notes that critical slowing down occurs at both limiting points of the first-order phase transition, i.e., at $\langle n \rangle =n_{c2}$ in the low activity state and  at $\langle n \rangle =n_{c1}$(or $n_{c1}(\alpha)$) in the high activity state.
Hysteresis was observed, for example, in living neural networks \cite{Soriano} and in simulations of thalamocortical systems \cite{izhikevich08}.

\subsection{Critical slowing down of neuronal dynamics}
\label{r-and-psd}

For deeper understanding of the first-order phase transition, we now find analytically the relaxation rate to the low activity state.
Writing Eq.~(\ref{steady-2}) in the form $\partial \Psi/\partial \rho_e +\partial \Psi/\partial \rho_i =1$ and substituting it into Eq.~(\ref{eigenv}), we find that at $\langle n\rangle=n_{c2}$ the eigenvalue $\lambda_{+}(\rho^{(1)})$ is zero at the fixed point 1. Therefore, the relaxation rate, Eq.~(\ref{r-rate}), to the low activity state is also zero,
\begin{equation}
\gamma_r =- \lambda_{+}(\rho^{(1)})=0.
\label{csd}
\end{equation}
This phenomenon is so-called critical slowing down. Note that it takes place on the line $\langle n\rangle=n_{c2}$ at all $\alpha$, both above and below $ \alpha_{t}$ (see Fig.~\ref{fig-overview}).

We now find dependence of the relaxation rate $\gamma_r$ on $\langle n\rangle$ at $0<n_{c2} - \langle n\rangle \ll n_{c2}$. We use the Taylor expansion of $\lambda_{+}$ over small $\delta \rho \equiv \rho^{(1)}(n_{c2})- \rho \ll \rho^{(1)}(n_{c2})$ in Eq.~(\ref{eigenv}) and obtain
\begin{equation}
\lambda_{+}(\rho)=\lambda_{+}(\rho^{(1)}(n_{c2}))+\frac{d\lambda_{+}(\rho)}{d\rho}\delta \rho +\dots .
\label{t-exp}
\end{equation}
The first term is zero. Using Eq.~(\ref{A-2}) for $\delta \rho $ from the Appendix \ref{app:r-rate}, in the leading order, we obtain
\begin{equation}
\gamma_r = -\lambda_{+}(\rho) \propto (n_{c2}-\langle n\rangle)^{1/2}.
\label{r-rate-1}
\end{equation}
This behavior occurs both at $\alpha >\alpha_t$ and $\alpha < \alpha_t$.

If a neuronal network has a finite but large size $N\gg 1$, then according to the scaling law hypothesis, near a critical point $n_{c}$ of a continuous  phase transition, the relaxation rate $\gamma_r$ is described by the general scaling law,
\begin{equation}
\gamma_r(\langle n \rangle,N) =(\langle n \rangle -n_c)^{\sigma}X[(\langle n \rangle -n_c) N^{1/\nu}]
\label{scaling law}
\end{equation}
with exponents $\sigma$ and $\nu$ which can be found by use of renormalization group techniques \cite{StaufferAharony,Stanley_book,Binder_1987}.  One assumes that the scaling law also is valid near the limiting point $n_{c2}$ of the first-order phase transition \cite{Sethna_2001}:
\begin{eqnarray}
\gamma_r(\langle n \rangle,N) &\propto & (\langle n \rangle /n_c {-}1)^\sigma, \,\,\text{if} \,\,\,\, N^{-1/\nu} \ll \langle n \rangle /n_c {-}1 \ll 1 \nonumber \\
&\propto & N^{-\sigma/\nu}, \,\,\,\,\,\,\, \text{if} \,\,\,\,\,\, \langle n \rangle /n_c -1 \ll N^{-1/\nu}.
\label{sl-2}
\end{eqnarray}
where $\sigma=1/2$.
Thus, at a finite but large size $N\gg 1$, the relaxation rate $\gamma_r$ is nonzero at any $\langle n \rangle$ due to finite size effects that smear the critical singularity. This agrees with results of our simulations.


\subsection{Power spectral density of fluctuations near the first-order phase transition}
\label{r-and-psd}

\begin{figure}
\includegraphics[width=0.33 \textwidth]{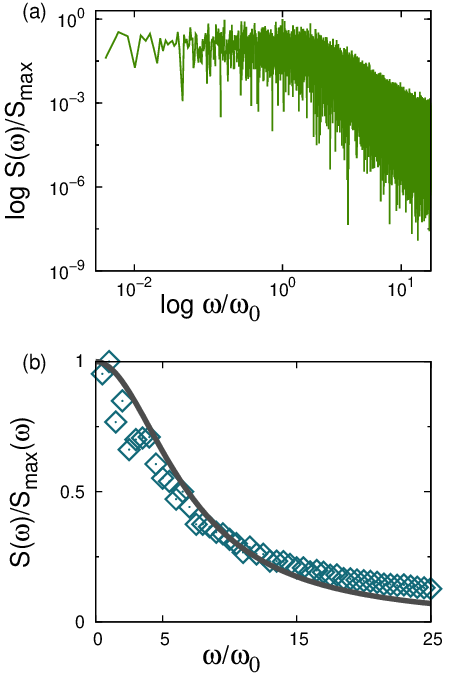}
\caption{(Color online)
(a) Power spectral density $S(\omega)$ of activity fluctuations versus frequency $\omega$ in the low activity state
observed in our simulations of the cortical
model at the shot noise intensity $\langle n\rangle = 18.7$ near the critical point of the first-order phase transition $n_{c2} \approx 18.8$ and $\alpha=0.85$ (region Ic).
(b). Averaged frequency dependence of $S(\omega)$ at small frequencies.
Results of simulations are shown by open diamonds. The solid line represents Eq.~(\ref{lmsp-PSDs}) with $\gamma_r = 6.9(2)$ and $\omega_0=0.03$. Frequencies are in units of $\mu_e$.
\label{fig-PSD-1}}
\end{figure}

We now find the power spectral density (PSD) of activity fluctuations
in the low activity state when $\langle n \rangle$ is close to $n_{c2}$.
In simulations, we measured the PSD
of excitatory and inhibitory activities.
We also solved analytically Eqs.~(\ref{eq:10})
with weak white-noise forces $F_e(t)$ and $F_i(t)$, where $F_e(t),F_i(t) \propto 1/\sqrt{N}$. The forces mimic forces caused by finite-size effects (this method was also used in \cite{bh1999}). Our calculations are  represented in Appendix \ref{PSD}.
We find that, in the low activity state, the PSD defined as
\begin{equation}
S(\omega) \equiv \langle \delta\rho_{e}(\omega) \delta\rho_{e}(-\omega)\rangle,
\label{PSD-1s}
\end{equation}
where $\delta\rho_{e}(t)\equiv\rho_{e}(t)-\rho^{(1)}$,
has a sharp zero frequency peak described by the following shape function (see Eq.~(\ref{sPSD-8})):
\begin{equation}
\frac{S(\omega)}{S_{max}}\approx \frac{1}{(\omega/\gamma_r)^2+1}.
\label{lmsp-PSDs}
\end{equation}
The peak maximum is $S_{max} \propto 1/ \gamma_{r}^2$. Fig.~\ref{fig-PSD-1}(a) displays the PSD $S(\omega)$ measured in our+ simulations in the low activity state in region Ic.
In Fig.~\ref{fig-PSD-1}(b) we compare simulations with the theoretical prediction. One sees that Eq.~(\ref{lmsp-PSDs}) describes well the measured frequency dependence of the PSD. According to Eq.~(\ref{r-rate-1}),  at $\langle n\rangle \rightarrow  n_{c2}$, the peak maximum increases as
\begin{equation}
S_{max} \propto 1/( n_{c2}-\langle n\rangle).
\label{PSD-2}
\end{equation}
Our simulations support the predicted increase of the zero frequency peak $S_{max}$ when $\langle n\rangle\rightarrow n_{c2}$. This behavior occurs if $\langle n\rangle$ is not very close to $ n_{c2}$. Due to finite-size effects,
$\gamma_r$ remains nonzero, though very small, even at the critical point (see Eq.~(\ref{sl-2})).

The Lorentzian behavior of the PSD of synaptic currents has been observed in cat cortex during wakefulness \cite{Bedard_2006a}. In Ref. \cite{Bedard_2006a}, it was suggested that this behavior  may be driven by a white noise process.
During slow-wave sleep, the PSD deviates from the Lorentzian \cite{Bedard_2006a}. This deviation suggests that, in a general case, stochastic forces may be statistically different from white noise.

Thus, the cortical model shows that bursts and avalanches
appear near the limiting point of metastable states of the first-order phase transition  caused by a saddle-node bifurcation in agreement with other network models \cite{Sethna_1993,Sethna_2001,mmkn10,Benayoun_2010}.
Critical phenomena (power-law statistics for avalanches and sharp zero-frequency peak of the PSD) due to critical slowing down in the low activity state (when approaching the critical point from below), the absence of the critical phenomena above the point (because, in the high activity state, the relaxation rate  is non-zero at the critical point), and hysteresis are the characteristic properties of the first order phase transition, which can be experimentally tested.
Another mechanism of avalanches based on ideas of self-organized criticality by Per Bak \cite{Bak} is discussed in \cite{Chialvo_2006}.
From our point of view, at the present time, there is no direct experimental evidence
that supports one approach over the other.
Further experimental and theoretical investigations
of these two approaches are necessary for understanding avalanches in the brain.

\section{Second-order non-equilibrium phase transitions}
\label{Non-equilibrium phase transitions}

We now consider the case $\alpha < \alpha_t$, i.e., when excitatory neurons respond faster on stimuli compared to inhibitory neurons. We show that, when increasing the shot noise intensity, the cortical model undergoes  successively two second-order phase transitions. We find that sustained network oscillations emerge at a saddle-node bifurcation and disappear at a Hopf bifurcation.  We study properties of the phase transitions, critical phenomena, patterns of spontaneous activity, and sustained network oscillations near the critical intensities of shot noise.

\begin{figure}
\includegraphics[width=0.45\textwidth]{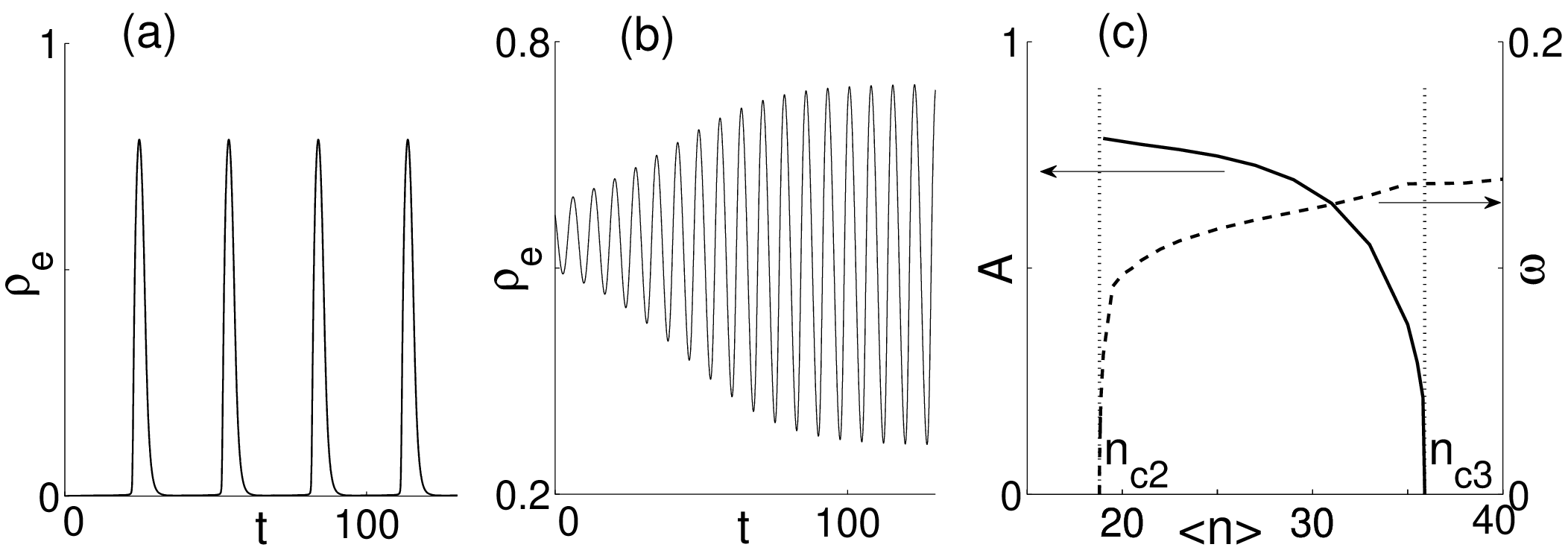}
\caption{
Network oscillations near (a) the saddle-node  ($\langle n \rangle=18.805, n_{c2}=18.8$) and (b) supercritical Hopf ($\langle n \rangle=34, n_{c3}=36$) bifurcations.
(c) Amplitude (solid line) and frequency (dashed line) of network oscillations versus
$\langle n \rangle$.
At $\langle n \rangle >n_{c3}$, the oscillations are damped. These results are obtained from a numerical integration of Eqs.~(\ref{eq:10}).
Time $t$ is in units of $1/\mu_e$, and $\alpha=0.75$.
\label{Hopf}}
\end{figure}

\subsection{Saddle-node bifurcation}
\label{Saddle-node bifurcation}

At a given $\alpha < \alpha_t$, we increase shot noise intensity $\langle n \rangle$ from $\langle n \rangle = 0$ (see line 2 in Fig.~\ref{fig-overview}). The neuronal network goes from region Ia with the single fixed point 1 into region Id or Ie where its dynamics
is determined by three fixed points: the stable point 1, the saddle point 2, and the unstable point 3 (see Table \ref{table1}).
At $\langle n \rangle = n_{c2}$ the points 1 and 2 coalesce and the network undergoes a second-order phase transition due to a saddle-node bifurcation from a state with a low activity and short-range temporal correlations between neurons into a state with regular sustained network oscillations (regions IIIa or IIIb).
In regions IIIa and IIIb, dynamics of neuronal networks is determined by the unstable fixed point 3 surrounded by a limit cycle.
At $\langle n\rangle >n_{c2}$,
the network oscillations emerge with a large amplitude (see Fig.~\ref{Hopf}(a)) and their frequency increases from zero as $\omega \propto (\langle n \rangle - n_{c2})^{1/2}$  (see Fig.~\ref{Hopf}(c)). This frequency dependence is typical for the saddle-node bifurcation in nonlinear dynamic equations \cite{Strogatz_book1994,Rinzel1989,Izhikevich2000}. Note however, that in our model, we deal with a phase transition, i.e., a collective phenomenon in neuronal networks. We suggest that for this kind of continuous phase transition the frequency is the order parameter.

\begin{figure}
\includegraphics[width=0.4\textwidth]{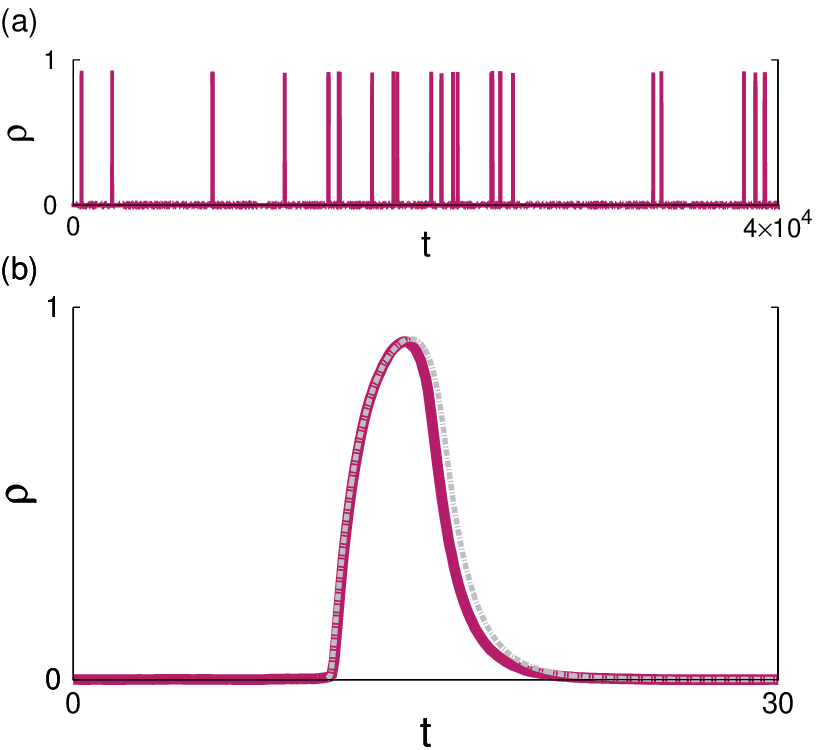}
\caption{(Color online) (a) Series of sharp spikes of neuronal activity near the saddle-node bifurcation.
(b) Paroxysmal-like spike of activity. Solid and dashed lines represent spikes found in simulations and numerical integration of Eqs.~(\ref{eq:10}), respectively. Parameters: noise intensity $\langle n \rangle=18.76$ and $\alpha=0.55$. Time $t$ is in units of $1/\mu_e$.
\label{sharpwave}
}
\end{figure}

In simulations, at $\langle n \rangle$ below $n_{c2}$, we observed  irregular almost identical sharp spikes of neuronal activity (see Fig.~\ref{sharpwave}(a)).
The mean frequency of the spikes is very small
and increases when the shot noise intensity tends to the critical point  $n_{c2}$ while the spike duration is almost constant and much larger than the period ($1/f$) of oscillations generated by a single neuron.
This kind of activity differs sharply from bursts found near the first-order phase transition (compare Figs. \ref{fig-hysteresis} and \ref{sharpwave}).
The sharp spikes emerge from a low background activity with a rapid onset (Fig.~\ref{sharpwave}(b)). They reach a large amplitude, involve in synchronized activity about 90 $\%$ of neurons, and end up with an abrupt return to lower activity. In Fig.~\ref{sharpwave}, the spike duration is about 0.2 s and the mean interspike interval is about 34 s at $1/\mu_e =20$ ms.

In order to understand the mechanism of generation of the sharp spikes, we performed numerical integration of Eqs.~(\ref{eq:10}) with non-zero stochastic forces $F_e$ and $F_i$ representing finite-size effects at the same parameters as in simulations.
The numerical integration also reveals sharp spikes that are identical to ones observed in simulations (Fig.~\ref{sharpwave}(b)).
Our analysis of the phase portrait of
Eqs.~(\ref{eq:10}) in regions Id and Ie shows that the sharp spikes are strongly nonlinear events in neuronal activity generated by fluctuations. In the ($\rho_i, \rho_e)$ phase plane, their trajectories
are topologically equivalent to the heteroclinic orbits
found in the Morris-Lecar model (see Fig. 7.4 in Ref. \cite{Rinzel1989}).

At $\langle n\rangle$ near $n_{c2}$,
the relaxation rate $\gamma_r$ is given by Eq.~(\ref{r-rate-1}). This result is in contrast to the standard mean field theory (the Landau theory) that predicts
$\gamma_r\propto |n_{c2}-\langle n\rangle|$  for a second-order phase transition. The non-standard scaling behavior and emergence of paroxysmal-like spikes near the saddle-node bifurcation
show an unusual character of the phase transition.

Analyzing properties of the sharp spikes, such as emergence conditions, course of the events, their shape, amplitude, duration, and low frequency oscillations, we find that this kind of spontaneous neuronal activity is similar to such epileptiform activity as the paroxysmal spikes
observed in EEG activity \cite{Steriade_1995,Timofeev_2004}.
Based on this similarity we suggest that the paroxysmal spikes
and other seizure-like events, such as slow-wave oscillations \cite{Timofeev_2004} or sharp waves in hippocampus \cite{Buzsaki_1985,Buzsaki_2006}, are possible strongly nonlinear waves appearing in neuronal networks near a saddle-node bifurcation. Of course, in order to describe in detail the events, a realistic network structure and realistic single-neuron dynamics must be taken into account.
As far as we know, paroxysmal-like spikes as collective nonlinear objects were not studied  analytically within a neuronal network model.
A detailed investigation of the nature and mechanism of generation of the paroxysmal-like spikes will be published elsewhere.

\subsection{Supercritical Hopf bifurcation}
\label{Supercritical Hopf bifurcation}

We now study the second-order phase transition due to the supercritical Hopf bifurcation. For this purpose we perform simulations of the cortical model, numerical integration, and analytical analysis of Eqs.~(\ref{eq:10}).
We find critical behavior and
demonstrate a difference in critical properties between the saddle-node and supercritical Hopf bifurcations.

\begin{figure}
\includegraphics[width=0.4\textwidth]{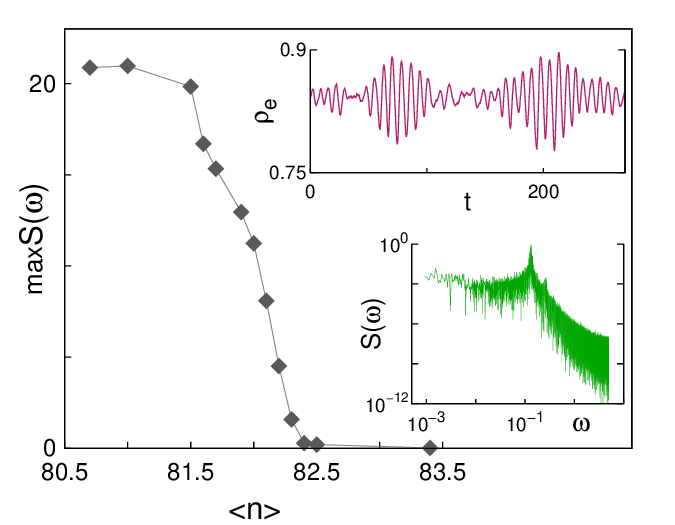}
\caption{(Color online)
The peak maximum $\max S(\omega)$ of the power spectral density (PSD)
of fluctuations versus the noise intensity $\langle n \rangle$ above the supercritical
Hopf bifurcation ($\langle n \rangle > n_{c3}\approx 80.5$).
Inset: Temporal neuronal activity in the form of spindles and the PSD $S(\omega)$ versus the frequency $\omega$ at $\langle n \rangle=82.5$. The simulations were performed at $\alpha=0.55$.
Time $t$ is in units of $1/\mu_e$.
\label{spindles}}
\end{figure}

When increasing the shot noise intensity $\langle n\rangle$ above $n_{c2}$, the frequency of sustained oscillations increases while their amplitude decreases (see Fig.~\ref{Hopf}(c)). The oscillations disappear at a critical noise intensity $\langle n \rangle =n_{c3}$ which depends on $\alpha$ (see the line 2 in Fig.~\ref{fig-overview}).
At $\langle n \rangle =n_{c3}$, the network undergoes a
phase transition from a state with the unstable point 3 surrounded by a limit cycle (region IIIa) into a state in which the fixed point 3 is a stable spiral (region IIb). From the stability analysis in Sec. \ref{steady states}, it follows that this transition is due to the supercritical Hopf bifurcation. Above $n_{c3}$ the network enters region IIb with damped network oscillations about the fixed point 3. Note also that network oscillations taking place near the saddle-node and supercritical Hopf bifurcations have different patterns (compare Figs.~\ref{Hopf}(a) and \ref{Hopf}(b)).
Oscillations emerging due to a Hopf bifurcation were also found in a stochastic rate model \cite{Wallace_2011}.

First, we study sustained network oscillations at $\langle n \rangle$ below $n_{c3}$. We expand  $\Psi(\rho_e,\rho_i)$ in Eqs.~(\ref{eq:10}) in a series in $\delta\rho_{a}(t)\equiv\rho_{a}(t)-\rho^{(3)}$ around the fixed point 3 and hold terms up to $O(\delta\rho_{a}^{3})$ inclusively.
Then, we solve Eqs.~(\ref{eq:10}) in region IIIa,
using the averaging theory \cite{Strogatz_book1994}. Details of our calculations are in Appendix ~\ref{oscillations-Hopf}.
When $\langle n \rangle \rightarrow n_{c3}$, we find a decrease of the oscillation amplitude $A$,
\begin{equation}
A \propto (n_{c3}  - \langle n \rangle)^{1/2},
\label{Hopf-1}
\end{equation}
and a decrease of the relaxation rate $\gamma_r$,
\begin{equation}
\gamma_r \propto |n_{c3}  - \langle n \rangle|,
\label{Hopf-2}
\end{equation}
that signals the supercritical Hopf bifurcation (see Fig.~\ref{Hopf}(c)). This behavior is typical for this kind of bifurcation \cite{Strogatz_book1994}. The amplitude $A$
is the order parameter for the transition. In Appendix ~\ref{oscillations-Hopf}, we show that a phase lag $\Delta\varphi$ between synchronized activities of excitatory and inhibitory populations is proportional to $\gamma_r$, $\Delta\varphi \approx \psi |\gamma_r|$. $\Delta\varphi$ is zero at $\langle n \rangle=n_{c3}$.
Comparing Eq.~(\ref{Hopf-2}) with Eq.~(\ref{r-rate-1}), we conclude that the continuous phase transitions corresponding to the saddle-node and supercritical Hopf bifurcations have different critical behaviors and, therefore, belong to different classes of universality.

\begin{figure}
\includegraphics[width=0.4\textwidth]{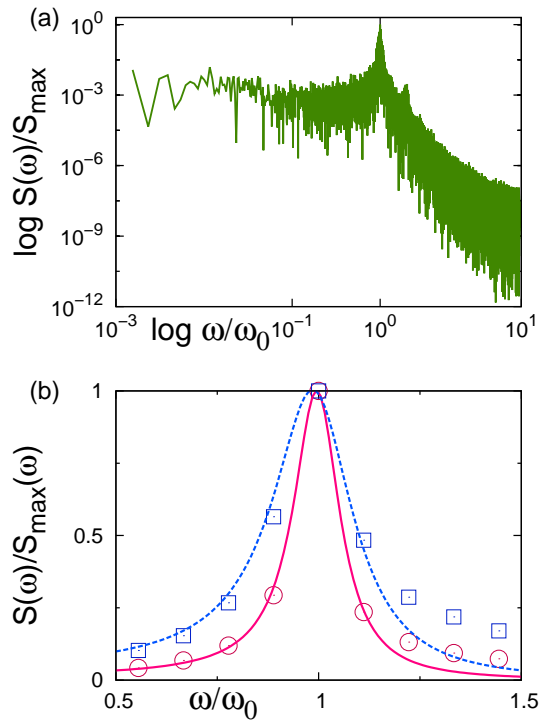}
\caption{(Color online)
Upper panel. Power spectral density $S(\omega)$ of activity fluctuations above the supercritical Hopf bifurcation observed in our simulation  at $\alpha=0.55$.
Lower panel. Averaged frequency dependence of $S(\omega)$ around the peak:
blue open rectangles represent simulation results at $\langle n \rangle=85$;
the blue dashed line represents the analytical calculation from Eq.~(\ref{PSD-3}) with $\gamma_r = 0.12(1)$, $\omega_0 = 0.15$;
pink open circles represent simulation results with $\langle n \rangle=82.5$;
the pink solid line represents our analytical calculation from Eq.~(\ref{PSD-3}) with $\gamma_r = 0.069(4)$, $\omega_0 = 0.15$.
Frequencies are in units of $\mu_e$.
\label{peakPSD}}
\end{figure}

We now analyze analytically the critical behavior of the cortical model at $\langle n \rangle$ above $n_{c3}$. Our simulations show that, above $n_{c3}$, spontaneous activity has a form of spindle oscillations
(see the inset in Fig. \ref{spindles}). The spindle oscillations are similar to spindles observed, for example, in thalamus \cite{cdss1996}. Damped oscillations were observed in an instance of epilepsy (see, for example, \cite{Babloyantz_1986}).
If $\langle n\rangle$ tends to $n_{c3}$ from above,
then the amplitude of spindle oscillations increases. This results in an increase of the peak of the power spectral density of activity fluctuations at the frequency of damped oscillations (see Fig. \ref{spindles}). This critical phenomenon signals an approach to the Hopf bifurcation.
In order to understand the phenomenon, we use simulations and analytical calculations.
According to Appendix \ref{psd-do}, the PSD, $S(\omega)$,
has a resonance peak,
\begin{equation}
S(\omega)/S_{max}\approx 4\zeta^2/[(1-x^2)^2+4\zeta^2 x^2],
\label{PSD-3}
\end{equation}
where $x=\omega/\omega_0$ and $\omega_0 \equiv \sqrt{\gamma_{r}^2 +\gamma_{i}^2}$. The parameter $\zeta$,
\begin{equation}
\zeta \equiv \gamma_r /\omega_0,
\label{zeta}
\end{equation}
is the damping ratio of damped oscillations. The peak maximum is $\max S(\omega) \equiv S_{max}\propto 1/\zeta^2$. This behavior of $S_{max}$ is due to the fact that the amplitude of damped oscillations increases as $A \propto 1/\zeta \propto 1/(\langle n \rangle - n_{c3})$ when $\langle n \rangle \rightarrow n_{c3}$ (see Eq. (\ref{correl-4})).
In turn, the amplitude decreases when $\langle n \rangle$ increases and the network goes away from the supercritical Hopf bifurcation.
Note that the relaxation rate $\gamma_r$ determines the time decay of the damped oscillations and can be found from data analysis of a time dependence of the autocorrelation function, Eqs.~(\ref{sPSD-2}) and (\ref{correl-4}). From this analysis one finds the dimensionless parameter $\zeta$ that is an important characteristic of the closeness of the network to the critical point $n_{c3}$. The smaller is $\zeta$ the close is the network to the critical point. In the infinite size limit, $\zeta$ is zero at $\langle n \rangle = n_{c3}$.
A similar resonance peak of the PSD was also found within the integrate-and-fire model in \cite{b2000,obh2009,lb2011}.

In Fig. \ref{peakPSD} we represent the PSD of activity fluctuations measured in simulations.
In agreement with the theoretical prediction, the measured PSD, $S(\omega)$,
reveals a sharp maximum at the frequency of damped oscillations. Fig.~\ref{spindles} shows that, when $\langle n \rangle \rightarrow n_{c3}$, the maximum value $S_{max}$ first strongly increases
and then saturates at a certain value due to the finite-size effects, Eq.~(\ref{sl-2}).
Fig.~\ref{peakPSD} shows that the shape of this maximum is well described by the shape function Eq.~(\ref{PSD-3}).

The critical behavior of the cortical model near the supercritical Hopf bifurcation helps to understand the attenuation of alpha rhythms  by visual or auditory stimuli  (the Berger effect) \cite{Hari_1997,Lehtel_1997}.
Recall that the Berger effect manifests itself in activation of alpha waves on the electroencephalogram when the eyes are closed and diminution of alpha waves when they are opened (see, for example, the review of \cite{Hari_1997}). Based on the cortical model, we suggest that opening eyes may result in an increase of the flow of spikes bombarding neurons in the area of the cortex that is responsible for the alpha waves. As a result, the neuronal network goes away from the Hopf bifurcation and the amplitude of damped oscillations decreases. A similar phenomenon was also observed in  the auditory cortex where the tau rhythm (the tau rhythm belongs to the family of alpha rhythms) was transiently suppressed by auditory stimuli \cite{Lehtel_1997}.

\section{Discussion}
\label{discussion}

In Sec.~\ref{Non-equilibrium phase transitions}, we discussed local stability of fixed points and bifurcations of nonlinear Eqs.~(\ref{eq:10}) in the cortical model in dependence on the shot noise intensity and the parameter $\alpha$.
Based on these results, one builds phase portraits of Eqs.~(\ref{eq:10}). In the case $\alpha < \alpha_t$, we revealed
that the phase portraits in regions Id, Ie, IIIa, and IIIb are topologically equivalent (in other words, homeomorphic) to the phase portraits found in the Morris-Lecar model stimulated by an applied current in the case when the $I{-}V$ relation is N-shaped \cite{Rinzel1989}.
Recall that the Morris-Lecar model is a simplified version of the four-dimensional Hodgkin-Huxley model.  Within the Morris-Lecar model, a system of two nonlinear equations  describes a relationship between the membrane potential and the activation of $K^{+}$ ion channels within the membrane.
It is well-known that the topological equivalence of phase portraits of two dynamical systems results in similar dynamics and similar responses on stimuli \cite{Kuznetsov_1998}.
Therefore, the dynamic behavior of the cortical model stimulated by shot noise (a flow of random spikes bombarding neurons) must be similar in some respects to the dynamic behavior of the Morris-Lecar model stimulated by an applied current.
In this case, we can apply results obtained for the well-studied Morris-Lecar model to the cortical model. Izhikevich \cite{Izhikevich2000} showed that the Morris-Lecar neuron acts as an `integrator', when it is close to the saddle-node bifurcation, and as a `resonator', when it is close to the Hopf bifurcation. Based on the topological equivalence, we can conclude that the cortical model acts in a similar way near the bifurcations. Indeed, in Sec. \ref{Saddle-node bifurcation},  we showed that if the mean frequency of incoming random spikes is a little bit larger than the critical frequency corresponding to the saddle-node bifurcation, then a neuronal network oscillates with an arbitrary low frequency. The higher the mean frequency of incoming random spikes, the higher the frequency of sustained network oscillations. Thus, we can say that the network acts as an `integrator'. In contrast, when the network is in the rest state near the supercritical Hopf bifurcation,  it acts as a `resonator' because it responds preferentially to a certain (resonant) frequency of input (see Sec. \ref{Supercritical Hopf bifurcation}). Furthermore, in Sec. \ref{Saddle-node bifurcation}, the topological equivalence helped us to understand the nature of paroxysmal-like spikes observed near the saddle-node bifurcation because similar nonlinear spikes were found in  the Morris-Lecar model by Rinzel and Ermentrout \cite{Rinzel1989}.

\section{Conclusion}
\label{Conclusion}

In conclusion, within an exactly solvable cortical model of neuronal networks with stochastic excitatory and inhibitory neurons, we studied first- and second-order phase transitions stimulated by shot noise (a flow of random spikes bombarding neurons). We performed
simulations, numerical integration, and analytical analysis of nonlinear dynamical equations. These methods gave results in complete agreement with each other.
The advantage of our model is that it gives a possibility to study both noise-induced first- and second-order phase transitions in neuronal networks by use of a unified approach and standard analytical physical and mathematical methods. This unified approach allowed us to compare qualitatively and quantitatively critical phenomena accompanying the phase transitions, patterns of spontaneous neuronal activity, and their dependence on the shot noise intensity. Furthermore, the rate equations derived for the model allowed us to study strongly nonlinear events, such as paroxysmal-like spikes and slow waves observed in neuronal activity, that cannot be described by a linear theory.
Our results support the idea that collective behavior of neuronal networks may have universal properties that do not depend on details of single neuron dynamics. The universal collective phenomena are determined by general properties of neuronal networks, such as the network structure, balance between
excitatory and inhibitory neurons, the presence of noise, and interaction between neurons.

We showed that if inhibitory neurons respond faster on stimuli than excitatory neurons, then a first-order phase transition manifests itself as a jump from low to high neuronal activity at a critical noise intensity. We found the mechanism of the transition and showed that it occurs due to a saddle-node bifurcation. We studied in detail critical phenomena that accompany the transition and patterns of spontaneous activity near the critical point. In particular, we showed that bursts and avalanches are precursors of the first-order phase transition.
When the shot noise intensity tends to the limiting point of the metastable states then the size distribution of neuronal avalanches becomes a power law with the exponent about 1.5.
Moreover, at the critical point, critical slowing down occurs in the infinite network, i.e., the relaxation rate is zero at the critical noise intensity. Our simulations revealed that finite-size effects smear the phase transition and make the relaxation rate to be non-zero at the critical point.
Critical phenomena (power-law statistics for avalanches and sharp zero-frequency peak of the PSD) due to critical slowing down in the low activity state (when approaching the critical point from below), the absence of the critical phenomena above the point (because, in the high activity state, the relaxation rate  is non-zero at the critical point), and hysteresis are the characteristic properties of the first order phase transition, which can be experimentally tested.

We also studied two noise-induced second-order phase transitions
that occur if inhibitory neurons respond slower on stimuli than excitatory neurons. These transitions represent two scenarios of appearance and disappearance of network oscillations.
When increasing the shot noise intensity, at first, sustained network oscillations appear due to a saddle-node bifurcation, and then, at a higher shot noise intensity, the oscillations disappear due to a supercritical Hopf bifurcation. Our analysis showed that the continuous phase transitions caused by the saddle-node and supercritical Hopf bifurcations are accompanied by different critical phenomena and different patterns of spontaneous neuronal activity. The transitions are characterized by different order parameters and belong to different classes of universality.

We analyzed patterns of spontaneous neuronal activity
near the saddle-node and Hopf bifurcations.
Our most interesting result is the observation of paroxysmal-like spikes that precede the second-order phase transition caused by the saddle-node bifurcation. We found that the spikes are strongly nonlinear objects
that appear instantly from a low background activity with a rapid onset, reaches a large amplitude, involve in synchronized activity about 90$\%$ of neurons,
and end up with an abrupt return to lower activity. These spikes are similar to single paroxysmal spikes and sharp waves observed in EEG measurements.
With increasing the shot noise intensity above the critical point of the saddle-node bifurcation, low frequency network oscillations follow the irregular spikes. They appear with a large amplitude but a small frequency (at the critical point, the frequency is zero).
The pattern of the oscillations resembles sharp-slow waves \cite{Timofeev_2004} or sharp waves in hippocampus \cite{Buzsaki_2006,Buzsaki_1985,Rex2009}.
In contrast to the saddle-node bifurcation, spontaneous activity above the Hopf bifurcation is represented by irregular spindles formed by damped oscillations.
Sustained network oscillations below the supercritical Hopf bifurcation have a small amplitude (at the critical point, the amplitude is zero in the infinite size limit) and a finite frequency. These oscillations are also nonlinear and have properties like ones of the Van der Pol oscillator.

We also analyzed the power spectral density (PSD) of spontaneous neuronal activity near the critical points of the phase transitions. We showed that the frequency dependence of the PSD and its dependence on the shot noise intensity give a rich information about the kind of bifurcation and the closeness of the network to the critical point. In particular, the PSD has a zero frequency peak near the first-order phase transition while above the supercritical Hopf bifurcation the PSD has a peak at the frequency of damped oscillations. The peaks are strongly enhanced when the noise intensity tends to the critical points of the phase transitions.
These results may be applied to an analysis of spectral properties of EEG recording in order to predict an approach to a critical point in neuronal activity.

Finally, we discussed an amazing similarity between excitability of the considered cortical model stimulated by shot noise and  excitability of the Morris-Lecar neuron stimulated by an applied current \cite{ML81,Rinzel1989,Izhikevich2000} . This similarity results from the fact that the cortical model of neuronal networks and the Morris-Lecar model have topologically equivalent phase portraits. This similarity allowed us to conclude that a neuronal network acts as `integrator', when it is close to the saddle-node bifurcation, and as a `resonator', when it is close to the supercritical Hopf bifurcation.
We believe this similarity may be useful for understanding many nonlinear phenomena in dynamics of neuronal networks.

In our model, a flow of random spikes bombarding neurons represents a combined effect of synaptic noise (spontaneous release of neurotransmitters), stimuli from other brain areas, and sensory stimuli. At given model parameters, the flow controls dynamics of the neuronal network. If the flow intensity is close to a critical value, then even a small change in the flow intensity can switch the network from one to another state. In other words, a small change of activity of neuronal networks to which a considered network is connected may strongly impact on a dynamic state of the network under consideration. This represents one of important mechanisms of interaction between neuronal networks \cite{Abbott_2006}.

\section{Acknowledgements}
We thank S. N. Dorogovtsev
for stimulating discussions.
This work was  supported by the PTDC Projects No.
SAU-NEU/ 103904/2008, No. FIS/ 108476 /2008, No. MAT/ 114515 /2009, the project PEst-C / CTM / LA0025 / 2011, and the project "New Strategies Applied to Neuropathological Disorders," cofunded by QREN and EU.
K.~E.~L. and M.~A.~L. were supported by the FCT Grants No. SFRH/ BPD/ 71883/2010 and No. SFRH/ BD/ 68743 /2010.

\appendix
\section{Neuronal activity near the critical points}
\label{app:r-rate}

Let us find the activity $\rho^{(1)}$ in the low activity state near the critical point $\langle n \rangle = n_{c2}$ of the saddle-node bifurcation, i.e., at $0< n_{c2}-\langle n \rangle \ll n_{c2}$. In simulations, $\rho^{(1)}$ can be found by measuring neuronal activity $\rho_e (t)$ and averaging it over a sufficiently large observation time. In Eq.~(\ref{steady-1}), we use the  Taylor expansion of the function $\Psi(\rho,\rho)$ over $\varepsilon\equiv \langle n \rangle -  n_{c2}$ and $\delta \rho \equiv  \rho - \rho^{(1)}(n_{c2})$ up to the second order in $\delta \rho$. Then Eq.~(\ref{steady-1}) takes a form,
\begin{equation}
\delta \rho = \frac{\partial \Psi}{\partial \langle n \rangle}\varepsilon{+}\frac{d \Psi}{d \rho} \delta \rho {+}\frac{1}{2} \frac{d^2 \Psi}{d \rho^2} (\delta \rho)^2 + ...,
\label{A-1}
\end{equation}
where the function $\Psi$ and its derivatives are calculated at $\langle n \rangle =n_{c2}$.
Using Eqs.~(\ref{steady-1}) and (\ref{steady-2}), we find a solution,
\begin{equation}
\delta \rho =\rho^{(1)} - \rho^{(1)}(n_{c2}) \approx - K ( n_{c2} - \langle n \rangle) ^{1/2}
\label{A-2}
\end{equation}
where
\begin{equation}
K=\Bigl|2\frac{\partial \Psi}{\partial \langle n \rangle} \Big/ \frac{d^2 \Psi}{d \rho^2}\Bigr|^{1/2}.
\label{A-2b}
\end{equation}
The singular behavior Eq.~(\ref{A-2}) is a general attribute of hybrid
and first-order phase transitions
\citep{Dorogovtsev_2006,Goltsev_2006,Baxter_2012}. Note that $\rho$ at the fixed point 2 is 
\begin{equation}
\rho^{(2)} - \rho^{(1)}(n_{c2}) \approx  K ( n_{c2} - \langle n \rangle) ^{1/2}
\label{A-2c}
\end{equation}
because, at $\langle n \rangle=n_{c2}$, the point 1 and 2 coalesce and $\rho^{(1)}(n_{c2})=\rho^{(2)}(n_{c2})$.

Neuronal activity $\rho^{(3)}$ near the Hopf bifurcation can be found at $0 < \langle n \rangle - n_{c3} \ll n_{c3}$ by use of the Taylor expansion Eq.~(\ref{A-1})
with $\varepsilon\equiv  \langle n \rangle - n_{c3}$ and $\delta \rho \equiv \rho - \rho^{(3)}(n_{c3})$,
where the function $\Psi$ and its derivatives are calculated at $\langle n \rangle =n_{c3}$. In this case, the linear terms give the leading contribution to a solution,
\begin{equation}
\rho^{(3)} - \rho^{(3)}(n_{c3}) \approx \frac{\partial \Psi}{\partial \langle n \rangle} \frac{\langle n \rangle - n_{c3}}{(1- d \Psi /d \rho)},
\label{A-4}
\end{equation}
in contrast to the square root dependence in Eq.~(\ref{A-2}).

\section{Power spectral density of activity fluctuations}
\label{PSD}

The power spectral density (PSD) of fluctuations of neuronal activity encodes rich information about critical phenomena.
According to the Wiener-Khintchine theorem, the power spectral density $S(\omega)$ of activity fluctuations of the excitatory population  is the Fourier transform of the autocorrelation function $C_{ee}(t)$,
\begin{equation}
S(\omega)=\frac{1}{2\pi}\int_{-\infty}^{\infty}e^{-i \omega t} C_{ee}(t) dt.
\label{sPSD-1}
\end{equation}
The autocorrelation function,
\begin{equation}
C_{ab}(t) \equiv
\frac{1}{T}\int_{0}^{T}\delta\rho_a(t_{1})\delta\rho_b(t_{1}+t) dt_1,
\label{sPSD-2}
\end{equation}
where  $\delta \rho_a(t)=\rho_a(t)-\rho_a$, describes fluctuations of activity $\rho_a(t)$ of population $a$, $a=e,i$, around the averaged value $\rho_a$.
$C_{ab}(t)$ is a measure of correlations between values of $\delta \rho_a (t_1)$ and $\delta \rho_b(t_1 +t)$ at two different instants separated by a lag $t$ and averaged over an arbitrary large time window $T$  (see, for example, in \cite{Gardiner_2002}).

In order to calculate the PSD, we assume that activity fluctuations are driven by weak white-noise forces $F_a(t)$  that mimic forces caused by finite-size effects,
\begin{equation}
\langle F_a(t)F_b(t')\rangle = F_{0}^2 \delta_{a,b}\delta(t-t')
\label{sPSD-3}
\end{equation}
where $F_{0} \propto 1/\sqrt{N}$. In this case, one can use the linear response theory  and find $\delta\rho_a(t)=\rho_a(t)-\rho$ from the linearized Eq.~(\ref{eq:10}),
\begin{equation}
\frac{d \overrightarrow{\delta\rho}}{dt}=(1-\rho)\overrightarrow{F} + \widehat{J}\overrightarrow{\delta\rho},
\label{sPSD-4}
\end{equation}
where $\rho$ is a steady state solution of Eq.~(\ref{steady-1}), $\overrightarrow{\delta\rho}=(\delta \rho_e,\delta \rho_i)$, and $\overrightarrow{F}=(F_e(t),F_i(t))$. The Jacobian $\widehat{J}$ is given by Eq.~(\ref{Jacobian}).
Making the Fourier transformation,
\begin{equation}
\delta \widetilde{\rho}_{a}(\omega) =\frac{1}{2\pi} \int_{-\infty}^{\infty} e^{-i \omega t} \delta \rho_{a}(t) d\omega,
\label{sPSD-5}
\end{equation}
we find the linear response,
\begin{eqnarray}
\delta \widetilde{\rho}_{e}(\omega)=\frac{(1-\rho)\Bigl[(i\omega-J_{22})\widetilde{F}_e(\omega)+J_{12}\widetilde{F}_i(\omega)\Bigr]}
{(i\omega+\lambda_{+})(i\omega+\lambda_{-})}, \nonumber \\
\delta \widetilde{\rho}_{i}(\omega)=\frac{(1-\rho)\Bigl[J_{21}\widetilde{F}_e(\omega)+(i\omega-J_{11})\widetilde{F}_i(\omega)\Bigr]}
{(i\omega+\lambda_{+})(i\omega+\lambda_{-})}.
\label{sPSD-6}
\end{eqnarray}
Substituting this result into Eq.~(\ref{sPSD-2}), we find the PSD for excitatory neurons,
\begin{equation}
S(\omega)=\frac{F_{0}^2 (1-\rho)^2}{2\pi}\frac{(J_{12}^2+J_{22}^2+ \omega^2)}{(\lambda_{+}^2+\omega^2)(\lambda_{-}^2+\omega^2)},
\label{sPSD-7}
\end{equation}
The PSD of inhibitory neurons is obtained from this equation after replacements: $J_{12} \rightarrow J_{21}$ and $J_{22} \rightarrow J_{11}$

\subsection{PSD near the first-order phase transition}

At first, let us consider the power spectral density (PSD) of activity fluctuations in the low activity state (the fixed point 1) in regions Ib and Ic in Fig.~\ref{fig-overview}. In these regions, eigenvalues $\lambda_{+}$ and $\lambda_{-}$ are real. When the noise intensity $\langle n \rangle$ tends to the critical point $n_{c2}$ of the
first-order phase transition, the eigenvalue $\lambda_{+}$ tends to zero according to Eq.~(\ref{r-rate-1}) while the eigenvalue $\lambda_{-}$ remains finite. Therefore, at small $\omega$, equation (\ref{sPSD-7}) takes a form
\begin{equation}
S(\omega)\approx \frac{F_{0}^2 (1-\rho)^2 (J_{12}^2+J_{22}^2)}{2\pi \lambda_{-}^2 \gamma_{r}^2}\frac{1}{(\omega/\gamma_{r})^2 +1},
\label{sPSD-8}
\end{equation}

\subsection{PSD near the Hopf bifurcation}
\label{psd-do}

Now we consider the PSD of activity fluctuations in the high activity state (the fixed point 3) at $\langle n \rangle > n_{c3}$ (region IIb in Fig.~\ref{fig-overview}). In this region the eigenvalues $\lambda_{\pm}$ are complex. Their real and imaginary parts determines the relaxation rate $\gamma_{r}$ and the frequency $\gamma_{i}$ of damped oscillations, respectively (see Eqs.~(\ref{r-rate}) and (\ref{d-freq})). In this case, equation (\ref{sPSD-7}) can be written in a form,
\begin{equation}
S(\omega)=\frac{F_{0}^2 (1-\rho)^2}{2\pi \omega_{0}^4}\frac{(J_{12}^2+J_{22}^2+ \omega_{0}^2 x^2)}{[(x^2 -1)^2+ 4\zeta^2 x^2]},
\label{sPSD-9}
\end{equation}
where $x\equiv \omega/ \omega_0$,
$\omega_0 \equiv [\gamma_{i}^{2} + \gamma_{r}^{2}]^{1/2}$, and $\zeta \equiv \gamma_{r}/\omega_0$. $\zeta$ is the damping ratio of the damped oscillations.

In the case when the shot noise intensity $\langle n \rangle$ tends from above to the critical point $n_{c3}$,
the relaxation rate $\gamma_r$ tends to zero (see Eq.~(\ref{Hopf-2})). If $\zeta\ll 1$, then the PSD has a sharp peak at the resonance frequency $\omega=\omega_r\equiv \omega_0 (1-2\zeta^2)^{1/2}$. The peak maximum is
\begin{equation}
S_{max}\equiv S(\omega_r)=\frac{F_{0}^2 (1-\rho)^2}{2\pi \omega_{0}^4}\frac{[J_{12}^2+J_{22}^2+ \omega_{0}^2 (1-2\zeta^2)]}{4 \zeta^2(1-\zeta^2)},
\label{sPSD-10}
\end{equation}
Near the resonance frequency $|\omega-\omega_r| \ll \omega_0$, $S(\omega)$ is described by a shape function $F(x,\zeta)$,
\begin{equation}
\frac{S(\omega)}{S_{max}}\approx F(x,\zeta)\equiv\frac{4\zeta^2 (1-\zeta^2)}{(1-x^2)^2+4\zeta^2 x^2}.
\label{correl-PSD}
\end{equation}

Substituting Eq.~(\ref{sPSD-6}) into Eq.~(\ref{sPSD-2}), we find that the autocorrelation function $C_{ee}(t)$ has a form,
\begin{equation}
C_{ee}(t)=A_e e^{-\gamma_r t}\cos\Big(\gamma_{i} t + \Phi_e\Big).
\label{correl-4}
\end{equation}
The amplitude $A_e$ and the phase  $\Psi_e$ behave as $A_e \propto 1 /\gamma_r$  and $\Phi_e \propto \gamma_r/\gamma_i$ at $\gamma_r \ll \gamma_i$. For inhibitory neurons we obtain a similar behavior.

\section{Oscillations near the supercritical Hopf bifurcation. nonlinear analysis.}
\label{oscillations-Hopf}

\begin{figure}
\includegraphics[width=0.45\textwidth]{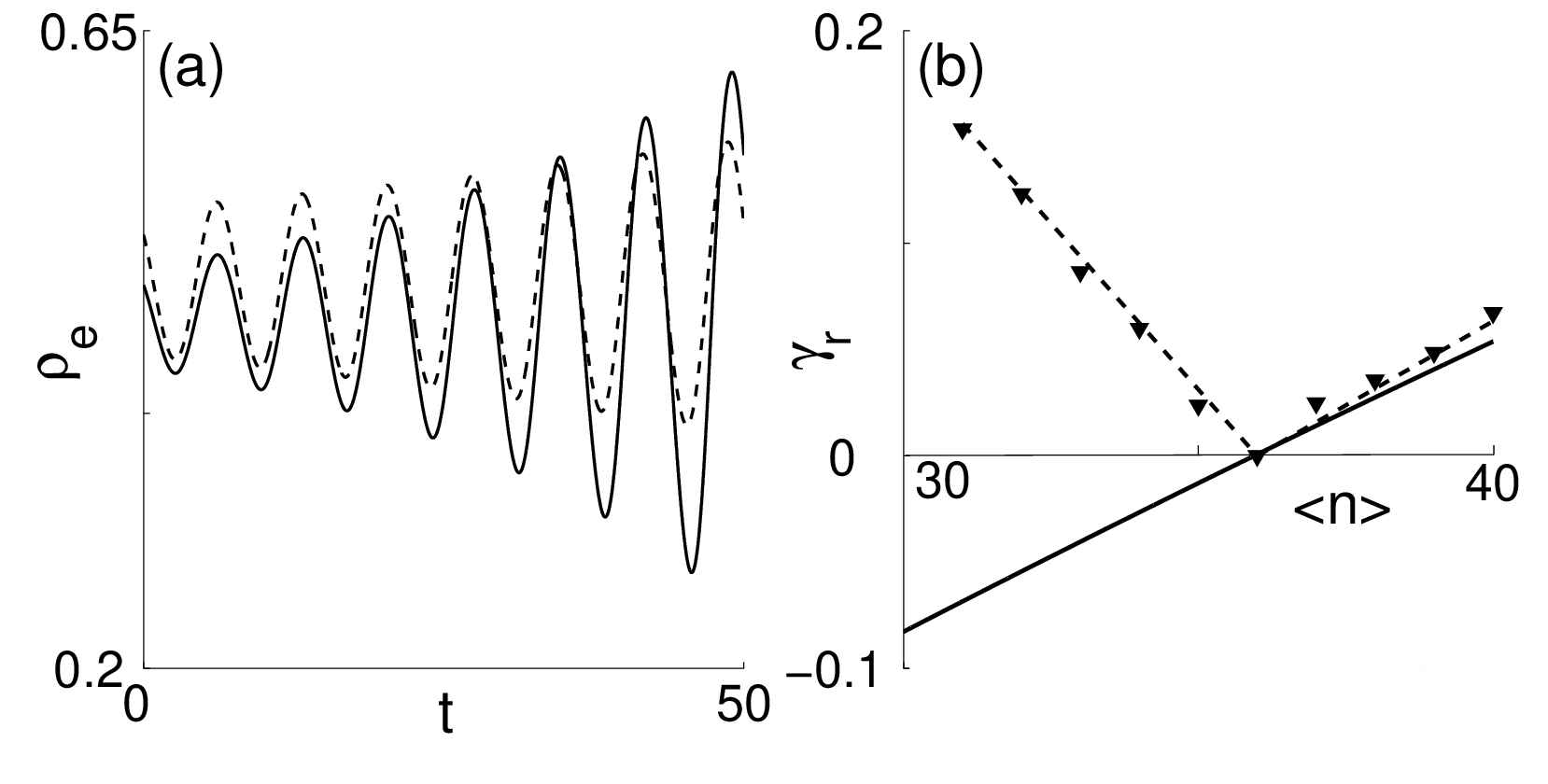}
\caption{(a) Relaxation dynamics of activity of excitatory neurons from an initial state to a state with sustained network oscillations at the shot noise intensity $\langle n \rangle $ below
the supercritical Hopf bifurcation:
(solid line) numerical integration of the approximate Eqs. (\ref{expansion});  (dashed line) exact Eqs. (\ref{eq:10}). (b) The parameter $\gamma_r$ (Eq.~(\ref{r-rate})) versus $\langle n \rangle $ from a numerical solution of Eq.~(\ref{steady-1}) at fixed point 3 (solid line).
The relaxation rate $\gamma_{r}^{*}$ obtained from a numerical analysis of
exact Eqs. (\ref{eq:10}) (dashed line).
Parameters $\langle n \rangle=35$, $\alpha=0.75$.
\label{hopf2}}
\end{figure}

In this appendix we study analytically
oscillations around the fixed point 3 near the Hopf bifurcation
in a noise intensity range $0< |n_{c3}-\langle n \rangle |\ll n_{c3}$ (a range around the boundary between regions IIIa  and IIb in Fig.~\ref{fig-overview}).
In this range, the oscillations
have a small amplitude that allows us
to use the Taylor expansion over $\delta\rho_a(t)=\rho_a(t) - \rho^{(3)}$ in  Eqs.~(\ref{eq:10}). Assuming $F_e=F_i=0$ and taking into account terms up to the third order in $\delta\rho_a(t)$, we obtain two coupled nonlinear equations,
\begin{widetext}
\begin{eqnarray}
&\frac{d \delta \rho_a(t)}{\mu_a dt} =-\delta\rho_a(t)+D^{(1,0)}\delta\rho_e(t)+D^{(0,1)}\delta\rho_i(t)
+\frac{1}{2}\big[D^{(2,0)}\delta\rho_e(t)^2+2D^{(1,1)}\delta\rho_e(t)\delta\rho_i(t)+D^{(0,2)}\delta\rho_{i}^{2}(t)\big] \nonumber \\
&+\frac{1}{6}\big[D^{(3,0)}\delta\rho_{e}^{3}(t)+3D^{(2,1)}\delta\rho_{e}^{2}(t)\delta\rho_i(t)
+3D^{(1,2)}\delta\rho_e(t)\delta\rho_{i}^{2}(t)+D^{(0,3)}\delta\rho_{i}^{3}(t)\big]
\label{expansion}
\end{eqnarray}
\end{widetext}
where $a=e,i$ and
\begin{equation}
D^{(n,m)}\equiv \frac{\partial^{n+m}\Psi}{\partial\rho_e^n\partial\rho_i^m}.
\label{D-terms}
\end{equation}
In Fig. \ref{hopf2}(a) we compare results of numerical integration of the reduced equations, Eqs.~ (\ref{expansion}), with
the exact equations (\ref{eq:10}). In the numerical integration, we studied relaxation of the system to a state with sustained oscillations (see Fig. \ref{hopf2}(a)) from an initial point $\rho_e =\rho_i = \rho^{(3)}$.
One sees that the frequency of the oscillations described by the reduced equations (\ref{expansion}) is very close to the frequency of oscillations from exact Eqs.~(\ref{eq:10}) though the amplitude of the sustained oscillations from Eqs.~(\ref{expansion}) is a little bit larger. These results evidence that the reduced equations (\ref{expansion}) are a good approximation to the exact Eqs.~(\ref{eq:10}). A similar analysis based on a reduced equation was used in \cite{bh1999,b2000} to study analytically oscillations near the Hopf bifurcation in networks of  integrate-and-fire neurons. Below we use  the reduced equations to study a critical behavior of the amplitude  of sustained oscillations, a relaxation rate to the state with the oscillations, and the phase lag between activities of excitatory and inhibitory populations.

It is convenient to rewrite Eqs.~(\ref{expansion}) in a vector form,
\begin{equation}
\delta\dot{\vec{\rho}}=\hat{J}\delta\vec{\rho}+\hat{M}(\delta \rho_e,\delta \rho_i)\delta\vec{\rho},
\label{vect-eq}
\end{equation}
where
\begin{displaymath}
\delta\vec{\rho} =
\left( \begin{array} {ccc}
\delta \rho_e \\
\delta \rho_i
\end{array} \right).
\end{displaymath}
$\hat{J}$ is the Jacobian Eq.~(\ref{Jacobian}) and $\hat{M}(\delta \rho_e,\delta \rho_i)$ is a matrix which introduces nonlinear terms,
\begin{eqnarray}
M_{11}&=&\frac{1}{2}D^{(2,0)}\delta\rho_e+\frac{1}{6}D^{(3,0)}\delta\rho_{e}^{2}, \nonumber \\
M_{12}&=&\frac{1}{2}D^{(0,2)}\delta\rho_i+\frac{1}{6}D^{(0,3)}\delta\rho_{i}^{2}+D^{(1,1)}\delta\rho_{e} \nonumber \\
&+&\frac{1}{2}D^{(2,1)}\delta\rho_{e}^{2} +\frac{1}{2}D^{(1,2)}\delta\rho_e\delta\rho_i,
\nonumber \\
M_{22}&=&\frac{\alpha}{2}D^{(0,2)}\delta\rho_i+\frac{\alpha}{6}D^{(0,3)}\delta\rho_{i}^{2}, \nonumber \\
M_{21}&=&\frac{\alpha}{2}D^{(2,0)}\delta\rho_e+\frac{\alpha}{6}D^{(3,0)}\delta\rho_{e}^{2}+\alpha D^{(1,1)}\delta\rho_{i} \nonumber \\
&+&\frac{\alpha}{2}D^{(1,2)}\delta\rho_{i}^{2} +\frac{\alpha}{2}D^{(2,1)}\delta\rho_e\delta\rho_i.
\label{matrix M}
\end{eqnarray}
The Jacobian $\hat{J}$, Eq.~(\ref{Jacobian}), can be represented in a form,
\begin{equation}
\hat{J}=-\gamma_r\hat{I}+(\vec{a}\hat{\vec{\sigma}})
\end{equation}
where $\hat{I}$ is the identity matrix. The parameter $\gamma_r$ is determined by Eqs.~(\ref{r-rate}) and (\ref{eigenv}) at the fixed point 3, i.e., $\rho_e=\rho_i=\rho^{(3)}$. In regions IIb and IIIa, Eq.~(\ref{eigenv}) gives $\gamma_r=(J_{11}+ J_{22})/2$. Furthermore,
$\vec{a}$ is a complex vector $\vec{a}{=}(a_1,a_2,a_3){=}\frac{1}{2}(J_{12}{+} J_{21},iJ_{12}{-}i J_{21},J_{11}{-} J_{22})$ with the property $\vec{a}^2=-\gamma_{i}^{2}$. We also use notations: $\hat{\vec{\sigma}}=(\hat{\sigma}_1,\hat{\sigma}_2,\hat{\sigma}_3)$ where $\hat{\sigma}_1,\hat{\sigma}_2$, and $\hat{\sigma}_3$ are the Pauli matrices.
Taking into account only linear terms in $\delta \rho_a$, the solution of Eq.~(\ref{vect-eq}) can be written in a form,
\begin{equation}
\delta\vec{\rho}=e^{-\gamma_r t+\vec{a}\hat{\vec{\sigma}}t}\vec{A} =
e^{-\gamma_r t} \big[ \cos(\gamma_{i}t)+\frac{\sin(\gamma_{i}t)}{\gamma_{i}}\vec{a}\hat{\vec{\sigma}}  \big ] \vec{A},
\label{l2}
\end{equation}
where the vector $\vec{A}=(A_e, A_i)$ is determined by an initial condition, $\vec{\rho}(t=0)=\vec{\rho}_0$. At $\langle n \rangle > n_{c3}$, i.e., in region IIb,  $\gamma_r$ is positive (see Fig. \ref{hopf2}(b)) and  the solution Eq.~(\ref{l2}) describes damped oscillations around the fixed point 3.

At $\langle n \rangle < n_{c3}$ in region IIIa, $\gamma_r$ is negative (see Fig. \ref{hopf2}(b)) and the solution Eq.~(\ref{l2}) is not valid due to the exponential divergence. In order to find a solution of Eq.~(\ref{vect-eq}), we must take into account the nonlinear terms.
In this case, we look for a solution in the following form:
\begin{equation}
\delta\vec{\rho}= e^{\vec{a}\hat{\sigma}t}\vec{A}(t).
\label{approx}
\end{equation}
Then Eq.~(\ref{vect-eq}) takes a form,
\begin{equation}
\dot{\vec{A}}=-\gamma_r\vec{A}+e^{-\vec{a}\hat{\sigma}t} \hat{M}(\delta \rho_e,\delta \rho_i)e^{\vec{a}\hat{\sigma}t} \vec{A},
\label{vect-eq-2}
\end{equation}
In the leading order in $\varepsilon=n_{c3} - \langle n \rangle $, in the limit $t \rightarrow\infty$, the oscillation amplitude $\vec{A}(t)$ tends to a stationary value that can be found by use of the averaging theory \cite{Strogatz_book1994}. We integrate Eq.~(\ref{vect-eq-2}) over the period $T=2\pi/ \gamma_i$ of oscillations,
\begin{equation}
0=\int_0^T \big[-\gamma_r\vec{A}+e^{-\vec{a}\hat{\sigma}t}\hat{M}(\delta \rho_{e}(t),\delta \rho_{i}(t))e^{\vec{a}\hat{\sigma}t}\vec{A} \big]dt,
\label{int-A}
\end{equation}
and obtain two coupled equations for $A_e$ and $A_i$,
\begin{eqnarray}
0&{=}& {-}\gamma_r A_e {+} a_{1}^{(e)}A_{e}^{3}{+}a_{2}^{(e)}A_{e}^{2}A_{i}{+}a_{3}^{(e)}A_{e}A_{i}^{2}{+}a_{4}^{(e)}A_{i}^{3}, \nonumber \\
0&{=}& {-}\gamma_r A_i {+} a_{1}^{(i)}A_{i}^{3}{+}a_{2}^{(i)}A_{i}^{2}A_{e}{+}a_{3}^{(i)}A_{i}A_{e}^{2}{+}a_{4}^{(i)}A_{e}^{3}.
\label{int-A-2}
\end{eqnarray}
where $a_{n}^{(a)}$ are coefficients. A simple analysis of these equations shows that, at $|\gamma_r| \ll 1$, a solution for the complex amplitudes $A_e$ and $A_i$ has a form,
\begin{equation}
\vec{A}=\sqrt{|\gamma_r|}\vec{B} \propto \sqrt{n_{c3} - \langle n \rangle}\vec{B},
\label{sq-A}
\end{equation}
where $\vec{B}=(e^{i\varphi_e} b_e, e^{i\varphi_i} b_i)$ is a complex vector and $\Delta\varphi \equiv \varphi_e -\varphi_i$ is a phase lag between excitatory and inhibitory activities.
The square root dependence in Eq.~(\ref{sq-A})
agrees with the numerical solution of Eqs.~(\ref{eq:10}) for the supercritical Hopf bifurcation (see Fig. \ref{Hopf}(c)). This dependence is a general property of the supercritical Hopf bifurcation.

Solving Eq.~(\ref{int-A}), we found that the phase lag $\Delta\varphi$ is proportional to $|\gamma_r|$,
\begin{equation}
\Delta\varphi \approx \psi |\gamma_r|.
\label{lag}
\end{equation}
The coefficient $\psi$ has a different value
above and below $n_{c3}$.
The phase lag $\Delta\varphi$ is zero at the critical point.

Equation (\ref{vect-eq-2}) also allows us to find the renormalized relaxation rate $\gamma_{r}^{*}$ of the neuronal activity to sustained nonlinear oscillations
with the equilibrium amplitude Eq.~(\ref{sq-A}). For this purpose, we look for a solution of Eq.~(\ref{vect-eq-2}) in a form,
\begin{equation}
\vec{A}(t)=(1-e^{-\gamma_{r}^{*} t})\vec{A}.
\label{approx}
\end{equation}
In agreement with results of numerical calculations displayed in Fig. \ref{hopf2}(b), we found
\begin{equation}
\gamma_{r}^{*} \approx G |n_{c3} - \langle n \rangle|,
\label{renorm-rate}
\end{equation}
where $G$  is a coefficient.



\end{document}